\documentclass[sigplan,screen]{acmart}

\copyrightyear{2024} 
\acmYear{2024} 
\setcopyright{rightsretained} 
\acmConference[ASPLOS '24]{29th ACM International Conference on Architectural Support for Programming Languages and Operating Systems, Volume 3}{April 27-May 1, 2024}{La Jolla, CA, USA}
\acmBooktitle{29th ACM International Conference on Architectural Support for Programming Languages and Operating Systems, Volume 3 (ASPLOS '24), April 27-May 1, 2024, La Jolla, CA, USA}
\acmDOI{10.1145/3620666.3651324}
\acmISBN{979-8-4007-0386-7/24/04}

\usepackage[]{hyperref}
\usepackage{amsmath,amsfonts}
\usepackage{graphicx}
\usepackage{textcomp}
\usepackage{xcolor}

\usepackage{multirow}
\usepackage{adjustbox}
\usepackage{caption}
\usepackage{subcaption}
\usepackage{subfiles}
\usepackage{tikz}
\usepackage{etoolbox}
\newrobustcmd*\circled[1]{\tikz[baseline=(char.base)]{
            \node[shape=circle,fill,inner sep=0.2pt] (char) {\textcolor{white}{#1}};}}

\usepackage{tabularx}
\usepackage{algorithm}
\usepackage[noend]{algpseudocode}
\usepackage{multirow}
\usepackage{array}
\usepackage{makecell}
\usepackage{xcolor, soul}
\sethlcolor{yellow}
\usepackage{xcolor}
\definecolor{commentcolor}{RGB}{102, 153, 0} % Define a new color for comment
\definecolor{olivegreen}{RGB}{0, 153, 0}
\usepackage{marginnote}
\usepackage{booktabs}
\usepackage{ifthen}
\newcommand{\algorithmcomment}[1]{\textcolor{commentcolor}{#1}} % Define a new command that typesets the argument in the comment color
\newcommand{\revision}{}

\newcommand{\rindex}[1]{}
\newcommand{\drindex}[1]{}

\newcommand{\revref}[1]{}
\newcommand{\figref}[1]{}
\newcommand{\tabref}[1]{}

\newcommand{\rspace}[1]{}
\begin{document}

%%
%% The "title" command has an optional parameter,
%% allowing the author to define a "short title" to be used in page headers.
\title{IANUS: 
 {I}ntegrated {A}ccelerator based on {N}PU-PIM {U}nified Memory {S}ystem}

\author{Minseok Seo} % removed for anonymity
%\email{sms0121@capp.snu.ac.kr}
\orcid{0000-0003-3648-5575}
\affiliation{%
\institution{Seoul National University}
%\city{Seoul}
\country{South Korea}
}
\author{Xuan Truong Nguyen} % removed for anonymity
%\email{sms0121@capp.snu.ac.kr}
\orcid{0000-0002-7527-6971}
\affiliation{%
\institution{Seoul National University}
%\city{Seoul}
\country{South Korea}
} 
\author{Seok Joong Hwang} % removed for anonymity
%\email{sms0121@capp.snu.ac.kr}
\orcid{0009-0008-2668-729X}
\affiliation{%
\institution{SAPEON Inc.}
%\city{Seoul}
\country{South Korea}
}

\author{Yongkee Kwon} % removed for anonymity
%\email{sms0121@capp.snu.ac.kr}
\orcid{0009-0001-9511-4734}
\affiliation{%
\institution{SK hynix}
%\city{Seoul}
\country{South Korea}
}
\author{Guhyun Kim} % removed for anonymity
%\email{sms0121@capp.snu.ac.kr}
\orcid{0000-0001-7076-2818}
\affiliation{%
\institution{SK hynix}
%\city{Seoul}
\country{South Korea}
}

\author{Chanwook Park} % removed for anonymity
%\email{sms0121@capp.snu.ac.kr}
\orcid{0009-0000-5764-5246}
\affiliation{%
\institution{SK hynix}
%\city{Seoul}
\country{South Korea}
}

\author{Ilkon Kim} % removed for anonymity
%\email{sms0121@capp.snu.ac.kr}
\orcid{0009-0009-0881-9292}
\affiliation{%
\institution{SK hynix}
%\city{Seoul}
\country{South Korea}
}

\author{Jaehan Park} % removed for anonymity
%\email{sms0121@capp.snu.ac.kr}
\orcid{0009-0003-9006-4904}
\affiliation{%
\institution{SK hynix}
%\city{Seoul}
\country{South Korea}
}

\author{Jeongbin Kim} % removed for anonymity
%\email{sms0121@capp.snu.ac.kr}
\orcid{0000-0002-5950-0056}
\affiliation{%
\institution{SK hynix}
%\city{Seoul}
\country{South Korea}
}

\author{Woojae Shin} % removed for anonymity
%\email{sms0121@capp.snu.ac.kr}
\orcid{0000-0001-9454-9966}
\affiliation{%
\institution{SK hynix}
%\city{Seoul}
\country{South Korea}
}

\author{Jongsoon Won} % removed for anonymity
%\email{sms0121@capp.snu.ac.kr}
\orcid{0009-0001-4632-0072}
\affiliation{%
\institution{SK hynix}
%\city{Seoul}
\country{South Korea}
}

\author{Haerang Choi} % removed for anonymity
%\email{sms0121@capp.snu.ac.kr}
\orcid{0000-0002-8933-6226}
\affiliation{%
\institution{SK hynix}
%\city{Seoul}
\country{South Korea}
}

\author{Kyuyoung Kim} % removed for anonymity
%\email{sms0121@capp.snu.ac.kr}
\orcid{0009-0005-9585-3835}
\affiliation{%
\institution{SK hynix}
%\city{Seoul}
\country{South Korea}
}

\author{Daehan Kwon} % removed for anonymity
%\email{sms0121@capp.snu.ac.kr}
\orcid{0000-0002-2033-8928}
\affiliation{%
\institution{SK hynix}
%\city{Seoul}
\country{South Korea}
}

\author{Chunseok Jeong} % removed for anonymity
%\email{sms0121@capp.snu.ac.kr}
\orcid{0009-0004-1666-7878}
\affiliation{%
\institution{SK hynix}
%\city{Seoul}
\country{South Korea}
}

\author{Sangheon Lee} % removed for anonymity
%\email{sms0121@capp.snu.ac.kr}
\orcid{0009-0006-9398-1507}
\affiliation{%
\institution{SAPEON Inc.}
%\city{Seoul}
\country{South Korea}
}

\author{Yongseok Choi} % removed for anonymity
%\email{sms0121@capp.snu.ac.kr}
\orcid{0009-0006-6512-0291}
\affiliation{%
\institution{SAPEON Inc.}
%\city{Seoul}
\country{South Korea}
}

\author{Wooseok Byun} % removed for anonymity
%\email{sms0121@capp.snu.ac.kr}
\orcid{0000-0002-2720-3102}
\affiliation{%
\institution{SAPEON Inc.}
%\city{Seoul}
\country{South Korea}
}

\author{Seungcheol Baek} % removed for anonymity
%\email{sms0121@capp.snu.ac.kr}
\orcid{0009-0007-1624-4586}
\affiliation{%
\institution{SAPEON Inc.}
%\city{Seoul}
\country{South Korea}
}

\author{Hyuk-Jae Lee} % removed for anonymity
%\email{sms0121@capp.snu.ac.kr}
\orcid{0000-0001-8895-9117}
\affiliation{%
\institution{Seoul National University}
%\city{Seoul}
\country{South Korea}
}

\author{John Kim} % removed for anonymity
%\email{sms0121@capp.snu.ac.kr}
\orcid{0000-0003-3958-3891}
\affiliation{%
\institution{KAIST}
%\city{Daejeon}
\country{South Korea}
}
%%% Should be placed below the authors
\renewcommand{\shortauthors}{Seo et al.}

\renewcommand{\arraystretch}{1.0}

\thanks{This paper is an updated version of the paper that appeared in the Proceedings of the 29th ACM International Conference on Architectural Support for Programming Languages and Operating Systems (ASPLOS), April 2024.} 

%%
%% The abstract is a short summary of the work to be presented in the
%% article.
\begin{abstract}
Accelerating end-to-end inference of transformer-based large language models (LLMs) is a critical component of AI services in datacenters. However, diverse compute characteristics of end-to-end LLM inference present challenges as previously proposed accelerators only address certain operations or stages (e.g., self-attention, generation stage, etc.). 
To address the unique challenges of accelerating end-to-end inference, we propose IANUS -- \textbf{I}ntegrated \textbf{A}ccelerator based on \textbf{N}PU-PIM \textbf{U}nified Memory \textbf{S}ystem. 
IANUS is a domain-specific system architecture that combines a Neural Processing Unit (NPU) with a Processing-in-Memory (PIM) to leverage both the NPU's high computation throughput and the PIM's high effective memory bandwidth.
In particular, IANUS employs a \emph{unified} main memory system where the PIM memory is used both for PIM operations and for NPU's main memory. 
The unified main memory system ensures that memory capacity is efficiently utilized and the movement of shared data between NPU and PIM is minimized.
However, it introduces new challenges since normal memory accesses and PIM computations cannot be performed simultaneously. 
Thus, we propose novel \emph{PIM Access Scheduling}
that manages
normal memory accesses and PIM computations through workload mapping and scheduling across the PIM and the NPU.
Our detailed simulation evaluations show that IANUS improves the performance of GPT-2 by 6.2$\times$ and 3.2$\times$, on average, compared to the NVIDIA A100 GPU and the state-of-the-art accelerator.
As a proof-of-concept, we develop a prototype of IANUS with a commercial PIM, NPU, and an FPGA-based PIM controller to demonstrate the feasibility of IANUS.

\end{abstract}

%%
%% The code below is generated by the tool at http://dl.acm.org/ccs.cfm.
%% Please copy and paste the code instead of the example below.
%%


%%
%% Keywords. The author(s) should pick words that accurately describe
%% the work being presented. Separate the keywords with commas.

%%
%% This command processes the author and affiliation and title
%% information and builds the first part of the formatted document.
\maketitle

%% Page numbers are excluded. - 240319
% \pagestyle{plain}

\section{Introduction}
Transformer \cite{vaswani2017attention}, BERT \cite{devlin2018bert}, and GPT \cite{radford2019language} have been widely used for natural language processing (NLP) services at datacenters.
Although GPUs are commonly used to accelerate the inference of deep learning models, GPUs are less effective in handling transformer models because of multi-head attention and inference stages that are memory-bound~\cite{hong2022dfx}. 
To address the limitations of GPU for transformer models, many recent works~\cite{ham20203,wang2021spatten,ham2021elsa,lu2021sanger} have proposed to accelerate multi-head attention through dedicated accelerators and algorithmic changes; however, these prior work do not fully address the challenges of end-to-end inference acceleration.
Recently, DFX~\cite{hong2022dfx} proposed an FPGA-based appliance that is designed for memory-bound transformer inference stages; however, it is sub-optimal for the compute-bound stages in end-to-end inference.

One of the main challenges in accelerating end-to-end inference of transformer-based large language models (LLMs) is their diverse characteristics, which exhibit a broad range of computational intensities.
For example, GPT includes complex vector operations, multi-head attention, and fully-connected (FC) layers that present both compute-bound matrix-matrix multiplication as well as memory-bound matrix-vector multiplication.
Consequently, to accelerate end-to-end inference of LLMs, hardware must be capable of efficiently handling 
all these diverse operations.

Neural processing units (NPUs)~\cite{chen2016eyeriss,isca-tpuv4,jouppi2017datacenter} have been widely proposed to accelerate deep neural networks (DNNs).
However, NPUs are often limited by memory-bound operations even when high-bandwidth memory is utilized. 
In comparison, processing-in-memory (PIM)~\cite{lee20221ynm,devaux2019true,kwon202125}  minimizes data movement by enabling computation near memory and provides higher effective memory bandwidth.
Recent
PIM chips \cite{lee20221ynm,kwon202125} are effective ``domain-specific'' memory as they accelerate memory-bound operations 
by guaranteeing full internal memory bandwidth utilization for processing units in memory on domain-specific kernels.
However, compute-bound operations such as matrix-matrix computations or complex vector operations are not efficient on PIM because of the limitations of  DRAM technology that is highly area-constrained.

To address the challenges of end-to-end LLM inference, 
we propose an NPU-PIM architecture that provides the benefit of both a domain-specific accelerator (i.e., NPU) as well as a domain-specific memory (i.e., PIM), effectively supporting a broad range of arithmetic intensities in LLMs. In particular, we propose 
IANUS --  \textbf{I}ntegrated \textbf{A}ccelerator based on \textbf{N}PU-PIM \textbf{U}nified Memory \textbf{S}ystem.~\footnote{IANUS is a Roman god with two faces that represented the middle ground between both concrete and abstract dualities.  The IANUS architecture in this work shares similarities as it represents a ``middle ground'' architecture between NPU and PIM architectures.} To the best of our knowledge, this is \textbf{\emph{one of the first works that integrate a commercial NPU with a commercial PIM memory to enable a domain-specific system architecture.}}
Previously proposed PIM-based systems view  PIM 
as an ``accelerator''~\cite{devaux2019true,kim2022aquabolt,kwon20221ynm,lee2021hardware} 
and employ a \emph{partitioned} memory system that uses the dedicated memory for the xPU (e.g., GPU, CPU) and the PIM accelerator memory. 
This leads to inefficient memory capacity usage as shared data between xPU and PIM tend to be duplicated in both memories for optimal performance.
This is especially problematic for LLM where parameters of FC layers represent a large portion of data that need to be shared between the NPU and the PIM.

\revision{In light of these challenges, we propose a \emph{unified} memory system where PIM memory also serves as the main memory for the NPU.}\drindex{R2b}
This approach removes the need for any data duplication and movement of shared data. 
However, the unified memory system in an NPU-PIM system introduces new challenges as PIM computations and normal memory accesses cannot be performed concurrently.
In this work, we propose a novel \emph{PIM Access Scheduling} (PAS) that 
schedules PIM computations and normal memory accesses through mapping and scheduling of the workload on the NPU-PIM architecture with a unified memory system.
The challenges of PIM computation in a unified memory system include memory resource conflict with normal memory accesses as well as the failure to leverage the potential for parallel execution with computations performed on the NPU.
Thus, PAS takes into account both resource conflicts and parallelizability of operations between the NPU and PIM to fully exploit the parallelism across the different resources.
We also demonstrate the proof-of-concept of IANUS by prototyping the system with an FPGA.
In summary, the key contributions of this work include the following.

\begin{enumerate}

\item \textit{Architecture}: We propose IANUS, a novel heterogeneous architecture that combines a dedicated hardware accelerator (NPU) with a specialized memory (PIM), to accelerate operations with diverse characteristics in the end-to-end LLM inference.

\item \drindex{R2c}\textit{Unified Memory System \& PIM Access Scheduling}: \revision{Identifying about 90\% of model parameters shared between the NPU and PIM in the LLM,} we propose a unified memory system 
where the memory for the NPU and the PIM memory is shared 
to efficiently utilize the memory capacity.
We also propose \emph{PIM Access Scheduling} (PAS) that manages the challenges of the unified memory system with effective workload mapping and scheduling. 
Through a detailed simulation of IANUS, IANUS with PAS achieves 6.2$\times$ and 3.2$\times$ speedup in GPT-2 compared to the A100 GPU and the state-of-the-art prior work (DFX~\cite{hong2022dfx}), respectively.

\item \textit{System integration and FPGA prototyping}: To demonstrate the feasibility of IANUS, we build an integrated system including a commercial NPU, commercial PIM chips, and an FPGA-based PIM controller.
\end{enumerate}
\section{Background}\label{background}

\begin{figure}[t]
\centering
\begin{subfloat}[][]
{\includegraphics[width=\linewidth]{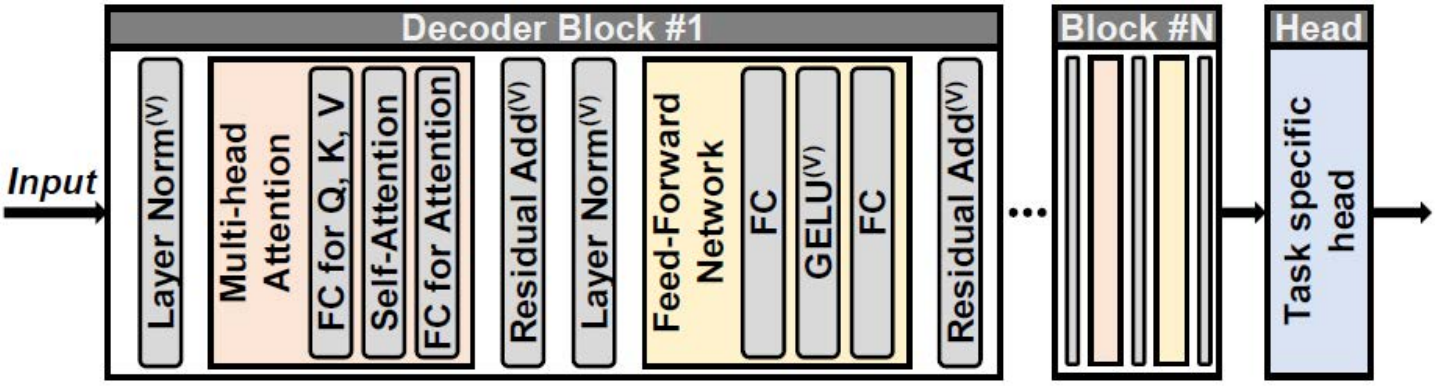}
\label{fig:GPT structure}}
\end{subfloat}
\begin{subfloat}[][]
{\includegraphics[width=\linewidth]{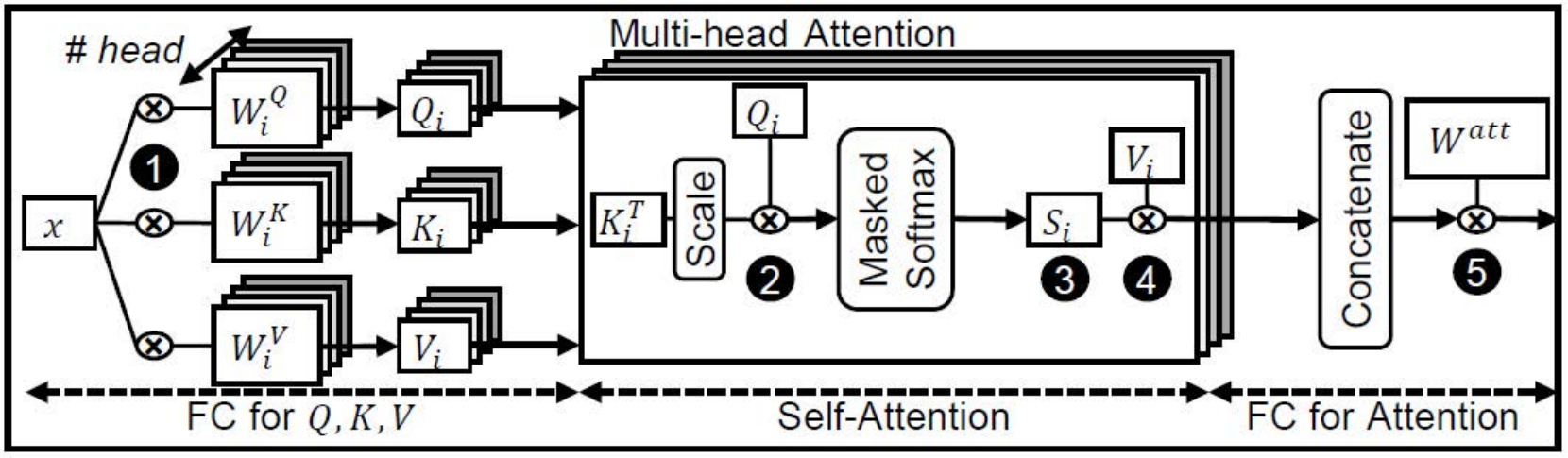}
\label{fig:Self attention}}
\end{subfloat}
\caption{(a) Structure of GPT with vector operations marked with \textsuperscript{(V)} and (b) multi-head attention mechanism shown in detail.}
\end{figure}

\subsection{Transformer-based LLMs}
NLP usually consists of two stages: input token summarization stage (\textit{summarization}) and output token generation stage (\textit{generation}).
While the \textit{summarization} stage
processes all input tokens collectively, the \textit{generation} stage deals with one generated token per stage.
In text generation tasks, the \textit{summarization} stage initially handles all inputs, followed by the \textit{generation} stage processing each produced token.

Transformer-based LLMs, such as BERT and GPT, use multiple encoder or decoder blocks, followed by a task-specific head (Figure \ref{fig:GPT structure}).
Each block consists of multi-head attention module, feed-forward network (FFN) module, layer normalization \cite{ba2016layer}, and residual addition \cite{he2016deep}.
During the \textit{summarization} stage, FC layers typically operate as matrix-matrix multiplication with multiple input tokens, while in the \textit{generation} stage, they perform matrix-vector multiplication with a single token.
The multi-head attention mechanism is depicted in Figure \ref{fig:Self attention}.
Input tokens ($x$) are multiplied with 
weight matrices to generate query ($Q$), key ($K$), and value ($V$) (\circled{1}).
In the \textit{generation} stage, new $K$ and $V$ are concatenated with previous ones.
For self-attention, $Q, K,$ and $V$ are splitted into multiple heads. The matrix product of query and transposed key ($QK^T$) (\circled{2}) are executed to compute attention score ($S$) (\circled{3}) and then output ($SV$) (\circled{4}) within each head is generated.
Finally, the outputs of all heads are merged and processed by the following FC layer (\circled{5}). 

\subsection{Platforms for DNN Inference}
\textbf{Domain-specific Accelerators}: 
DNN accelerators~\cite{albericio2016cnvlutin,chen2014diannao,hegde2018ucnn,kwon2018maeri,liu2015pudiannao,jouppi2017datacenter,norrie2021design} mainly focus on accelerating convolution computation.
Therefore, these accelerators often face bandwidth bottlenecks during the \textit{generation} stage of LLMs, primarily involving matrix-vector multiplication. To tackle this problem, DFX \cite{hong2022dfx}, an FPGA-based appliance, maximizes bandwidth utilization by designing peak FLOPS to match the memory bandwidth.  However, while providing significant benefits on the  \textit{generation} stage,  the benefits of DFX on the \textit{summarization} stage are small because of limited FLOPS. 

\textbf{Processing-in-Memory}:
PIM refers to the technology of implementing processing units inside memory to accelerate specific workloads or save energy consumption.
Recently, PIM based on commercial DRAMs have been announced (Accelerator-in-Memory (AiM) \cite{lee20221ynm,kwon20221ynm}, HBM-PIM \cite{kwon202125,lee2021hardware}, and UPMEM-PIM~\cite{devaux2019true}). 
They are suitable for memory-intensive workloads by utilizing all- or half-bank parallelism.
As a result, they are considered promising solutions for \textit{generation} stages of LLMs because LLMs include matrix-vector multiplication in \textit{generation} stage.

\section{Motivations}

In this section, we show the diverse computation requirements of LLMs and present challenges in designing the accelerator system for their end-to-end inference. This motivates the need for a heterogeneous architecture that combines a domain-specific \emph{accelerator} with high compute capability
and a domain-specific \emph{memory} with high memory bandwidth. 
We also demonstrate the motivation of a unified main memory organization in an NPU-PIM system for LLMs.

\begin{figure}[t]
\centering
\begin{subfloat}[][]
{
\includegraphics[width=\linewidth]{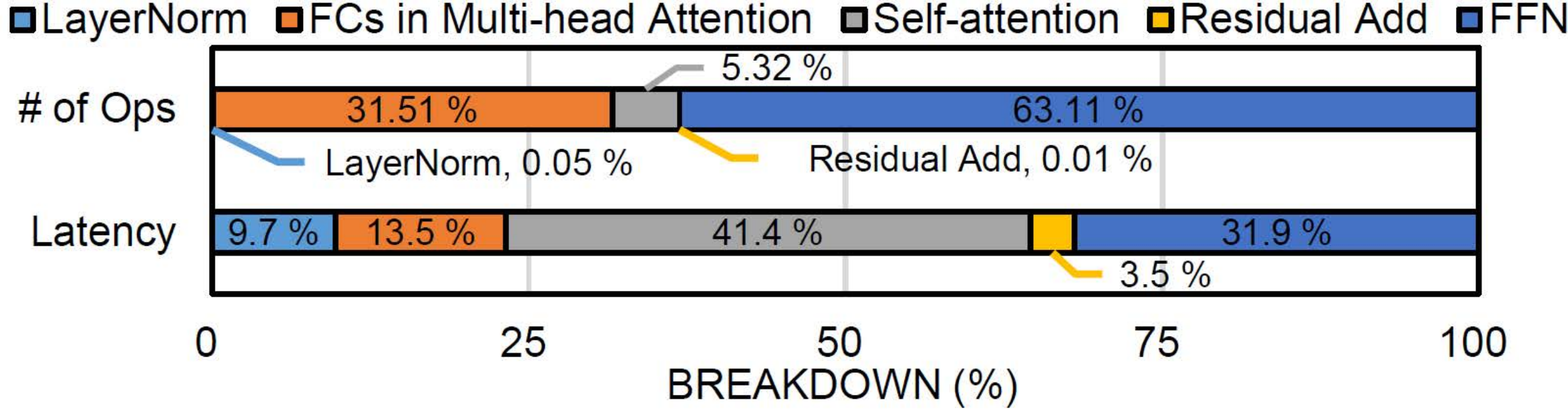}
\label{fig:encoder breakdown}
}
\end{subfloat}
\begin{subfloat}[][]
{
\includegraphics[width=\linewidth]{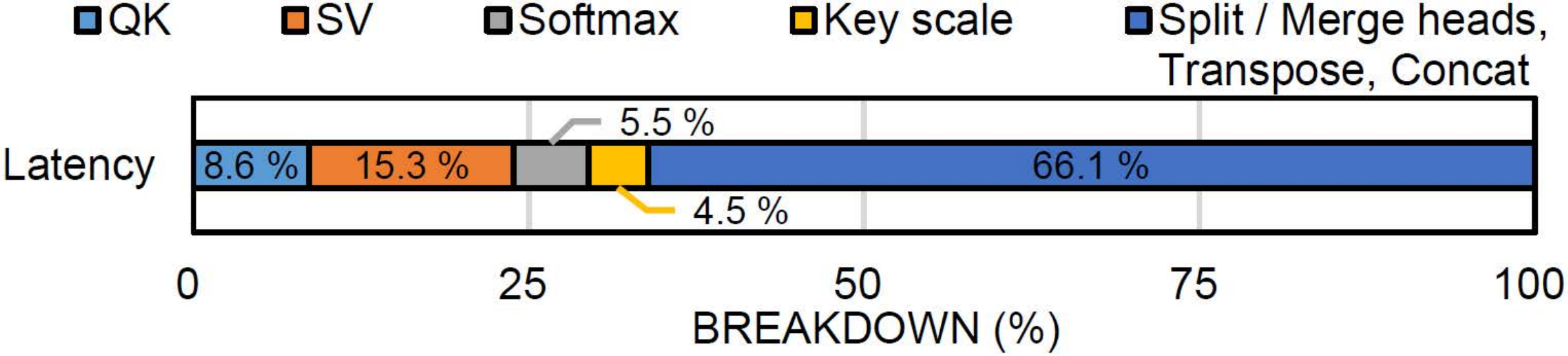}
\label{fig:attention breakdown}
}
\end{subfloat}
\caption{\textit{Generation} stage of GPT-2 XL (a) Latency and FLOPs breakdown of decoders. (b) Latency breakdown of self-attention. Results are obtained using an A100 GPU.}
\end{figure}

\subsection{Diverse Computational Requirements of LLMs}

The  \textit{generation} and the \textit{summarization} stages exhibit different computational characteristics as the \textit{generation} stage of LLMs is often memory-bound with matrix-vector operations while 
\textit{summarization} stage is compute-bound with matrix-matrix operations.
While compute-bound components are well-matched to compute accelerators (e.g., NPU, GPU), memory-bound operations are not.
For example, when generating two tokens with 512 input tokens, the \textit{generation} stage requires 512$\times$ fewer FLOPs compared to the \textit{summarization} stage.  
However, the execution time of the \textit{generation} stage is 88.5\% of the \textit{summarization} stage on A100 GPU.
As shown in Figure \ref{fig:encoder breakdown}, 
FCs and FFNs in the \textit{generation} stage that consist of matrix-vector multiplications account for 45.4\% of the total latency and 
are well-matched to be accelerated by PIM.
In comparison, the \textit{summarization} stage shows an even greater reliance on FCs and FFNs, which mainly employ matrix-matrix multiplications, thereby necessitating a compute accelerator (e.g., NPU) for effective acceleration.
Thus, in this work, we propose a heterogeneous accelerator that integrates both an NPU with a  PIM to address the diverse computation requirements in LLMs.

In addition, LLMs also include vector operations such as layer normalization and non-computing operations such as transpose of matrix. As shown in Figure \ref{fig:encoder breakdown}, layer normalization and residual addition represent 13.2\% of the total latency while representing less than 0.06\% of the total FLOPs, raising a need for a dedicated vector processing unit. Additionally, a significant portion of self-attention latency in the decoder is attributed to non-computing operations within the self-attention, as in Figures \ref{fig:encoder breakdown} and \ref{fig:attention breakdown}. Among operations in self-attention that accounts for 41.4\% of the total decoder latency, non-computing operations occupy 66.1\% of the total self-attention latency. This substantial impact of non-computing operations highlights the necessity for a domain-specific accelerator with flexible data manipulation.

\begin{figure*}[t]
\centering
\begin{center}
\includegraphics[page=3,clip, trim=0.065cm 8.45cm 0.1cm 1.82cm,width=0.97\linewidth]{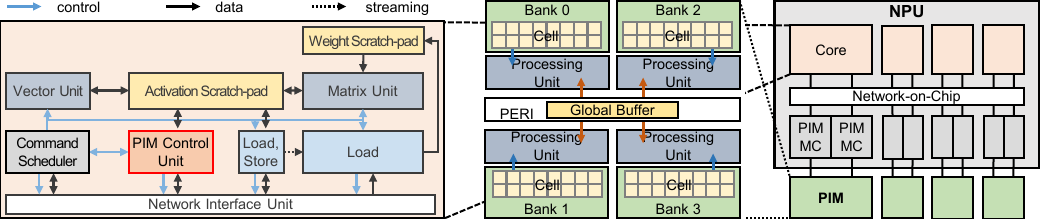}
\end{center}
\caption{(Left) Architecture of a core in NPU. (Middle) PIM architecture. (Right) Overall architecture of IANUS. }
\label{fig:IANUS}
\end{figure*}

\subsection{Partitioned vs. Unified Memory Systems in LLMs}
Systems using commercial PIM~\cite{devaux2019true,kim2022aquabolt,kwon20221ynm,lee2021hardware} with CPU or GPU typically employ a partitioned main memory system where some main memory is dedicated for PIM accelerator's memory while the remaining memory is used by the host (i.e., CPU or GPU).
This approach can maximize parallelism as both PIM and the host can access their own memory.  However, partitioned memory can be problematic if there is significant sharing of data between the host and the PIM accelerator as the same data need to be duplicated across both memories to maximize the parallelism.
Without duplicating data, substantial data transfers between two memories are necessary, potentially deteriorating performance.

In LLMs, the parameters of FC layers need to be shared between the NPU and the PIM since they are utilized both in the matrix-matrix and matrix-vector computation.
Since the FC parameters constitute a large fraction of data required for inference (e.g., 91\% in GPT-2), 
using a partitioned memory in the NPU-PIM system for LLMs results in inefficient usage of the memory.
As a result, we employ a unified memory organization where the PIM is used as the main memory for both the PIM accelerator and the NPU -- resulting in approximately 2$\times$ reduction in memory footprint compared to partitioned memory system. 

However, a unified memory presents new challenges, compared to the partitioned memory system, as the PIM memory is responsible for both ``normal'' memory accesses from the NPU as well as the PIM computation and these two steps cannot be executed in parallel.
As na\"ive scheduling does not consider memory resource conflicts between PIM computations and normal memory accesses and fails to observe the parallelizability between PIM computations and other computations, it cannot exploit available parallelism across the NPU and the PIM.
In this work, we propose \emph{PIM Access Scheduling} that addresses such challenges of the unified memory system.

\section{IANUS Architecture}

To accelerate the end-to-end inference of transformer-based LLMs, we introduce IANUS (\textbf{I}ntegrated \textbf{A}ccelerator based on \textbf{N}PU-PIM \textbf{U}nified Memory \textbf{S}ystem) that integrates NPU and PIM (Figure \ref{fig:IANUS}).
This section describes the IANUS architecture, including the NPU and the PIM architecture that we leverage, details the transformer-aware microarchitecture within NPU and PIM, as well as introduces new microarchitectural components that we propose to enable IANUS with a unified memory system architecture.

\subsection{NPU \& PIM Architecture} \label{sec:comp}

\textbf{Computation Units in NPU:}
 As in Figure \ref{fig:IANUS}, a single core of NPU comprises two computing units: the matrix unit (MU) and the vector unit (VU). The MU is built on a systolic array \cite{kung1979systolic} of 128$\times$64 processing elements to accelerate matrix-matrix multiplication, such as FC layers. To enable efficient pre- or post-processing, the MU also supports output scaling and bias addition.
The VU consists of sixteen 4-wide VLIW processors \cite{fisher1983very}.
As it is designed to manage vector operations and general purpose operations that the MU cannot efficiently perform, the VU supports element-wise addition, layer normalization \cite{ba2016layer}, masking, and non-linear activation functions such as softmax \cite{bridle1989training} and GELU \cite{hendrycks2016gaussian}.

\textbf{Scratch-pad Memories in NPU:}
The activation scratch-pad memory (AM) and the weight scratch-pad memory (WM) in the core of NPU supply data to the computing units. The WM provides weights, scales, and biases to the matrix unit. The AM serves as a data storage for both computing units, typically providing input or activation data. The AM adopts a transposed data addressing layout relative to the WM to fully exploit the benefits of the matrix unit’s systolic array.
Moreover, the size of data accessed by a single address in each scratch-pad (\emph{entry}) is aligned with the corresponding dimension of the matrix unit's systolic array. Specifically, the entry size of the AM is twice that of the WM.

\textbf{PIM Architecture:}
PIM architecture for IANUS is based on the commercial PIM (AiM \cite{kwon20221ynm,lee20221ynm}) that i) exploits true all-bank parallelism, ii) is designed to accelerate end-to-end matrix-vector multiplication and activation functions in DRAM, and iii) is based on commodity DRAM (GDDR6).
Processing units (PUs) are implemented at each bank of the PIM
and a global buffer
is implemented at the peripheral circuit (Figure~\ref{fig:IANUS}). The global buffer is shared between all PUs and stores an input vector, often reused multiple times when processing matrix-vector products. In comparison, large data with low reusability such as weight matrix, often read just once during matrix-vector product, are stored at each bank. Each PU, associated with each bank, includes a set of multipliers, an adder tree, an accumulator for Multiply-Accumulate (MAC) operation, and an activation function unit.

\subsection{Transformer-Aware NPU \& PIM Microarchitecture}
In this subsection, we highlight the NPU microarchitecture designed to accelerate self-attention and vector operations in transformer-based LLMs, along with the data allocation scheme in PIM aimed at optimizing FC operations.
\subsubsection{Data Manipulation in Self-Attention}\label{Sec: attention}
\textbf{\\Key Transpose:}
The transpose operation requires data transfer between on-chip and off-chip memory without dedicated hardware, potentially delaying PIM operations that also utilize off-chip memory.
We avoid off-chip access by executing transpose within on-chip through incorporating a streaming path between DMAs (light blue boxes in Figure~\ref{fig:IANUS}) of two scratch-pads.
However, moving data from the activation scratch-pad (AM) to the weight scratch-pad (WM) through on-chip DMA only performs a partial transpose operation because of the mismatch of the entry sizes for the two scratch-pads. 
Thus, we introduce a streaming buffer between the two scratch-pads for on-chip data movement during on-chip DMA.
We then implement weight interleaving within the matrix unit, enabling access to the WM entry with a specific stride.

\textbf{Splitting / Merging Attention Heads:}
Splitting and merging attention heads represent a large fraction of the self-attention latency in a GPU due to the data reordering. Our compiler avoids such data reordering by carefully defining and generating activation scratch-pad addresses of input and output data in the command. For instance, when generating commands for the FC operation that produces $Q$, the compiler generates as many commands as the number of heads. The compiler then assigns a distinct output address for each command, guiding the matrix unit to store $Q$ in the scratch-pad in a split manner. Hence, no data reordering overhead is required. Similarly, the compiler ensures consecutive output addresses of each head's $SV$ command for merging attention heads.

\subsubsection{Vector Operations in Vector Unit}
\textbf{\\Layer Normalization:}
Given the limited amount of on-chip memory within the vector unit (VU), 
a two-phase approach is used where VU calculates the mean and variance of the tokens in the first phase while the normalization is done in the second phase.

\textbf{Masked Softmax:}
We combine masking and softmax \cite{bridle1989training} within a single kernel.
Each mask is stored as a 1-bit bitmap, reducing data movement and memory usage.
In softmax, we subtract the max value for stability instead of the large value.

\textbf{GELU:}
For the GELU activation \cite{hendrycks2016gaussian}, VU uses a lookup table (LUT) approximation,
widely employed due to its accuracy and performance \cite{hong2022dfx,yu2022nn}. GELU activation is also supported in PIM by reserving some DRAM rows inside PIM as LUT for the activation function and linearly interpolating data from the LUT within the processing unit of PIM.

\begin{figure}[t] %%% t: top, b: bottom, h: here
\begin{center}
\includegraphics[page=2,clip, trim=0.1cm 0.7cm 7.4cm 10.1cm,width=1.0\linewidth]{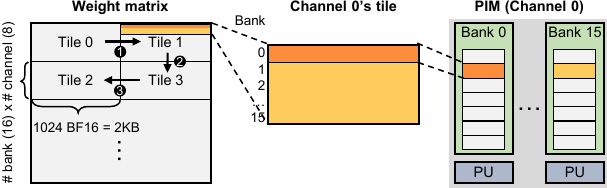}
\end{center}
\caption{Data allocation and tiling scheme for a matrix-vector multiplication in PIM.}
\label{fig:PIM_FC}
\end{figure}

\subsubsection{Data Allocation in PIM}
\textbf{\\}PIM is exploited for matrix-vector multiplication during the FC layers in the \textit{generation} stages.
We exploit data allocation and tiling scheme that maximize the performance of FC layers in PIM and demonstrate these strategies with the weight matrix of an FC layer in Figure \ref{fig:PIM_FC}.
The weight matrix is divided into tiles with each tile consisting of 16 (number of banks per channel) $\times$ 8 (number of channels for IANUS) rows and up to 1024 columns (number of elements in one DRAM row). Each row in the tile is allocated to the same DRAM row address across each bank and each channel to maximize performance as PIM can perform computation across all banks and all channels in parallel.
While the optimal tiling can vary across workloads, we assume row-major tiling.

\subsection{IANUS Microarchitecture}
\textbf{Command Scheduler:}
The command scheduler is responsible for checking dependencies between each command and the status of each compute, DMA, and PIM unit and sending commands to each unit.
When a command has no dependency and the corresponding unit
is in an idle state, the scheduler pushes the command into the “issue” queue of the unit, and the unit executes it. 
If the command cannot be issued, the command is pushed into the “pending” queue.
Upon completion of execution, the scheduler
resolves the dependencies between the command and the other commands.

\textbf{PIM Control Unit and PIM Memory Controller:}
Orchestrating multiple PIM chips is not trivial as it requires scheduling a large number of PIM commands and increases the complexity of the command scheduler.
More importantly, the efficiency of PIM computation diminishes if a standard memory command is inserted in the middle of multiple PIM commands for a single ``operation'', such as a matrix-vector multiplication.
Thus, we propose \emph{macro} PIM command for scheduling. One macro PIM command, which represents a single operation, comprises multiple \emph{micro} PIM commands (e.g., a single matrix-vector operation is executed through a macro PIM command that consists of multiple micro PIM commands, including providing the input vector, performing the MAC operation, etc.). To support macro PIM commands, a PIM control unit (PCU) and PIM memory controller (PIM MC) are added, as shown in Figure \ref{fig:IANUS}.

When one macro PIM command reaches ``ready'' state, the command scheduler forwards the macro PIM command to the PCU. At the same time, the scheduler puts other DMA commands related to the off-chip memory into ``wait'' state to ensure the PIM execution is not interrupted. Once the PCU receives the macro PIM command, PCU decodes it into multiple micro PIM commands and forwards these to the PIM MC through the network-on-chip (NoC).

The PIM MC supports both PIM commands and normal memory commands.
Similar to conventional memory controllers, PIM MC tracks the state of each memory bank and generates appropriate commands following pre-defined timing constraints as well as newly introduced states and timing constraints of PIM operations.
When all micro PIM commands within one macro PIM command finish, the completion signal is forwarded to the command scheduler to enable DMA commands associated with the off-chip memory.

\rspace{
\phantomsection~\label{sec: write traffic}
\revision{
Managing DMA and PIM commands via the command scheduler's blocking scheme not only guarantees uninterrupted PIM computation but also helps to mitigate issues caused by write traffic. In unified memory systems, data correctness and consistency issues often arise from the writing of shared data between NPU and PIM. However, pausing DMA commands during PIM executions prevents such data integrity issues. Furthermore, in the context of transformer inference on IANUS, write traffic issues are alleviated since the shared data – the parameters of FC layers – are read-only.
}
}

\textbf{Network-on-chip:} 
The NoC topology in IANUS provides all-to-all connectivity between all of the cores and the PIM MCs. The NoC traffic for IANUS consists of both the memory traffic as well as the PIM traffic to support the unified memory system. All-to-all connectivity ensures that each core can access any memory channel when PIM is used as the main memory of the NPU. In addition, the NoC is also used for PIM commands from the PCU to the PIM MCs. The NoC supports broadcasting of PIM commands to all PIM MCs to reduce NoC bandwidth demand while providing parallel PIM operations across all PIM channels.

\begin{figure}[t] %%% t: top, b: bottom, h: here
\begin{center}
\includegraphics[page=35,clip, trim=0cm 12.4cm 7.9cm 0.6cm,width=1.0\linewidth]{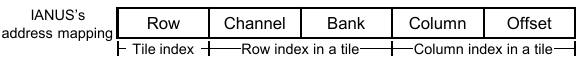}
\caption{IANUS's DRAM address mapping with the mapping of tile shown in Figure \ref{fig:PIM_FC}.}
\label{fig:address mapping}
\end{center}
\end{figure}

\textbf{DRAM Address Mapping:}
\rindex{R3a}\revision{
The DRAM address mapping of IANUS is shown in Figure~\ref{fig:address mapping}. 
IANUS employs an address mapping of (MSB) Row-Channel-Bank-Column (LSB) and the main goal of the IANUS's mapping is to maximize PIM computation performance through PIM-aware tile (Figure~\ref{fig:PIM_FC}) placement.
By using the row address bit as the MSB and using those bits as the index of a tile, data within a single tile share the same row address
that ensures row conflicts do not occur during the compute operations related to a single tile.
In addition, each tile is assigned to a different row address. 
The column address bit is used as the LSB to ensure that operations on all elements of a single row within a tile are handled by one processing unit to execute MAC within one bank.
Placing channel and bank address bits between the row and column address bits allows each row within a tile to be distributed across different channels and banks.
This enables the PIM to concurrently compute all rows within a tile by leveraging channel and bank parallelism and maximize PIM computation throughput.
}

\begin{figure}[t] %%% t: top, b: bottom, h: here
\begin{center}
\includegraphics[page=1,clip, trim=0.13cm 5.05cm 7.1cm 5cm,width=1.0\linewidth]{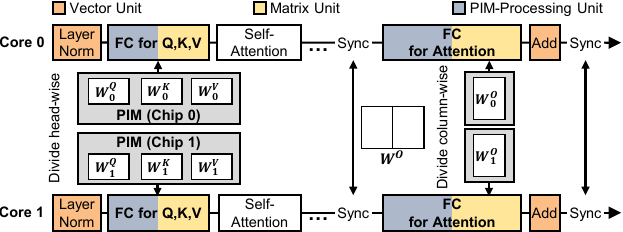}
\end{center}
\caption{Workload mapping and execution flow, featuring intra-layer parallelism and attention head parallelism. For simplicity, only one attention head per core is shown. The mapping of operations in self-attention is detailed in Section~\ref{schedule_section}.}
\label{fig:workload}
\end{figure}

\begin{algorithm}[t]
\caption{Adaptive mapping algorithm for FC layers.}
\label{Algo: workload}
\begin{flushleft}
\textbf{Input/Output:} \textit{CMDs} (a sequence of commands) \\
\textbf{Params:} \textit{n} (number of input tokens), \textit{T} (the size of MU)\\
\textbf{Define:} $VU, MU, PIM, DMA$ (analytical model of units)
\end{flushleft}
\begin{algorithmic}[1]
  \For{$i, cmd$ in $CMDs$}
    % \State $t_{prefetch} \gets 0$
    \If{$cmd.type == MU_{FC}$}
      \State $prev\_cmd \gets CMDs[i-1]$
      \State \algorithmcomment{// Check prefetching} 
      \If{$prev\_cmd.type == VU$}
        \State $t_{prefetch} \gets VU(n, prev\_cmd.dim)$
      \EndIf
      \State \algorithmcomment{// Consider tiling and pipelining for MU} 
      \State $w_{cfg} \gets cmd.weight\_cfg$
      \State $w_{load} \gets DMA_{weight}(w_{cfg})$
      \State $mu_{FC} \gets MU_{FC}(n, w_{cfg})$
      \State $mu_{time} \gets pipe((w_{load},mu_{FC}),T) - t_{prefetch}$
      
      \State \algorithmcomment{// Calculate PIM time} 
      \State $pim_{time} \gets n \times PIM(w_{cfg})$

      \If{$pim_{time} < mu_{time}$}
        \State Replace $CMDs[i].type$ with $PIM$
      \EndIf

    \EndIf

  \EndFor

\end{algorithmic}
\end{algorithm}

\section{PIM Access Scheduling}
\label{sect_pim_access_schedule}
The integrated NPU-PIM architecture with a unified main memory presents challenges as the main memory is used by both the NPU and the PIM compute logic. In this section, we propose \emph{PIM Access Scheduling} (PAS) that enables efficient sharing of the physical memory between NPU and PIM.
Unlike traditional memory access scheduling~\cite{rixner2000memory} that involves scheduling of memory commands, PAS not only needs to consider scheduling normal DRAM commands and PIM commands but also needs to address the challenges of workload mapping across the NPU and the PIM. More importantly, the scheduling or mapping of the workloads impacts how the DRAM/PIM commands are scheduled. In this section, we describe PAS within the context of IANUS, particularly on FC operations and multi-head attention in LLMs, and how they are mapped/scheduled on IANUS.

\subsection{Overview}~\label{sec: workload}
We present the execution flow and workload mapping of LLMs for IANUS in Figure \ref{fig:workload}.
To leverage parallelism across all cores in the NPU as well as across all PIM chips, we exploit attention head parallelism by partitioning the weights of the FC for $Q, K,$ and $V$ across PIM chips in a head-wise scheme.
Through the head-wise partitioning, each core can access the memory in parallel to load the weights or the output of PIM compute for the multi-head attention.

For other FC operations, we leverage intra-layer parallelism to minimize data movement of weights that are considerably larger than input or activation data in LLMs.
To reduce synchronization overhead between each core in the NPU, we partition the weights of FC column-wise.
Synchronization occurs four times: once after multi-head attention, twice after each residual addition, and once after GELU.
Meanwhile, layer normalization and residual addition are mapped to the vector unit (VU) within the NPU (Figure \ref{fig:workload}).

\phantomsection~\label{sec: fine-grained}
\subsection{FC Operation}
FC can be mapped to either the matrix unit (MU) or the PIM.
The \textit{summarization} stage often has a large input token size and results in high computation requirements -- thus, it is more appropriate to map the FC to the matrix unit.
When the input token size is small, loading the weights from the memory can become the bottleneck because the arithmetic intensity of the FC operation decreases.
Thus, an adaptive mapping algorithm is necessary to determine whether to map the FC to the PIM or the MU within the NPU. 

An overview of the adaptive mapping algorithm is summarized in Algorithm \ref{Algo: workload}.
To determine the appropriate unit for FC, we develop a simple analytical model that estimates the execution time across different execution units (e.g., MU, VU, DMA, PIM) based on the number of input tokens at compile time.
The input for the adaptive mapping algorithm is a sequence of commands based on mapping of the FC to the matrix unit. 
When estimating the time of FC on MU, we assume a pipelined scheme for both weight loading and computation, as well as tiling configured to match the MU's size (lines 8-11).
We also account for weight prefetching time if an operation of VU precedes the FC operation (lines 5-6).
We then compare the estimated time of FC on MU with that of PIM and assign the FC to the execution unit that can complete sooner (lines 13-15).
If the first FC of FFN is mapped to the PIM, the GELU will also be allocated to the PIM since the PIM is designed to support GELU right after FC.

\subsection{Multi-Head Attention}
\label{schedule_section}

\begin{figure}[t] %%% t: top, b: bottom, h: here
\centering
\begin{subfloat}[][]
{
\includegraphics[page=5,clip, trim=0.1cm 8.9cm 7.53cm 2.2cm,width=1.0\linewidth]{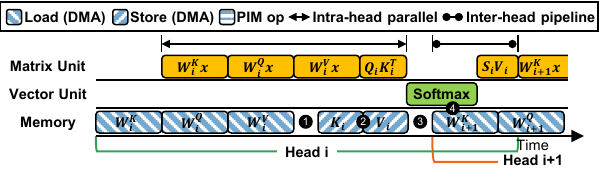}
\label{fig:attn_NPU}
}
\end{subfloat}
\begin{subfloat}[][]
{
\includegraphics[page=5,clip, trim=0.1cm 6.6cm 7.53cm 5.3cm,width=1.0\linewidth]{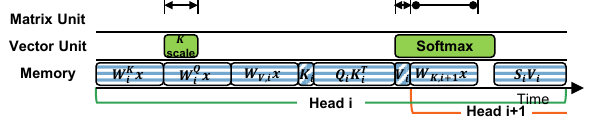}
\label{fig:attn_no_opt_PIM}
}
\end{subfloat}
% \\[-0.2ex]
\begin{subfloat}[][]
{
\includegraphics[page=5,clip, trim=0.1cm 4.4cm 7.53cm 7.4cm,width=1.0\linewidth]{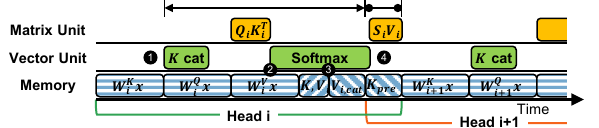}
\label{fig:attn_opt_PIM}
}
\end{subfloat}
\caption{Unified memory-aware scheduling for multi-head attention at (a) \textit{summarization} stage where FCs are mapped to the matrix unit and \textit{generation} stage where FCs are mapped to the PIM: $QK^T$ and $SV$ mapping to (b) PIM or (c) matrix unit. Figures (b) and (c) are drwan on the same time scale to show the latency difference.}
\end{figure}

As described earlier in Figure \ref{fig:Self attention}, multi-head attention consists of a series of operations that have various computational requirements. While IANUS provides computing capability of both NPU and PIM, na\"ive scheduling that overlooks parallelizability and resource conflicts between operations may lead to the under-utilization of both units with considerable latency overhead.
To address this challenge, we propose unified memory-aware scheduling for multi-head attention at both the \textit{summarization} and \textit{generation} stages.

\textbf{\textit{Summarization} stage:}
As shown in Figure \ref{fig:attn_NPU}, FC layers for $Q, K,$ and $V$ typically operate as matrix-matrix multiplications with multiple input tokens ($x$) -- thus are computed in the matrix unit, while weight matrices ($W^{Q,K,V}$) are loaded from the memory via DMA.
To efficiently process multi-head attention, we utilize both intra-attention head parallelism and inter-attention head pipelining.
We prioritize key generation to execute key transpose in parallel with value generation.
As DMAs are utilized for on-chip transpose, they are not used for PIM access during transpose (\circled{1}).
Given that the matrix unit supports output scaling (Section \ref{sec:comp}), the key scaling operation is omitted.
We also ensure that key and value are stored during computations (\circled{2}).
To hasten the start of the $SV$ operation, values are moved to the weight scratch-pad via on-chip data transfer during the softmax (\circled{3}).
In addition, we utilize inter-attention head pipelining by prefetching the weight of the next head (\circled{4}).

\textbf{\textit{Generation} stage:} FC layers mainly perform matrix-vector multiplications with one input token ($x$), making them well-suited for PIM computation.
Similarly, since $QK^T$ and $SV$ operations involve matrix-vector multiplications and require loading previously generated keys and values, their executions can appear to be more suitable for PIM.
As shown in Figure \ref{fig:attn_no_opt_PIM}, mapping $QK^T$ and $SV$ to PIM avoids such load operations.
However, the overall performance benefit is limited since parallelism across both the PIM and the NPU cannot be exploited well as the PIM performs most of the operation.
In addition, computing $QK^T$ and $SV$ in PIM results in poor efficiency because of the mismatch between the PIM DRAM row size and the data size.
For example, with a head dimension of 64, PIM computational efficiency of $QK^T$ is only 6.25\% due to only 64 BF16 elements being utilized for computation out of the 1024 elements available in one DRAM row.

As a result, we propose mapping $QK^T$ and $SV$ operations to the matrix unit, and accordingly, scheduling based on this mapping.
To exploit inter-attention head parallelism, as shown in Figure \ref{fig:attn_opt_PIM}, we execute key concatenation in the vector unit instead of storing the key (\circled{1}), enabling its simultaneous execution with query generation in PIM.
Loading the previously generated keys ($K_{pre}$) of $i$th head is omitted in Figure~\ref{fig:attn_opt_PIM}, as its small size compared to the FC weight allows for prefetching.
We then transpose concatenated keys within on-chip while performing query generation in PIM.
Furthermore, we execute $QK^T$ and softmax respectively in parallel with value generation by mapping $QK^T$ to matrix unit (\circled{2}).
After value generation, storing generated keys and values and loading concatenated values ($V_{cat}$) are performed during softmax (\circled{3}).
We also employ inter-attention head pipelining by prefetching $K_{pre}$ of the next head during $SV$ (\circled{4}).
If the prefetching ends before the completion of $SV$, the key generation of the next head is performed in conjunction with $SV$.
Consequently, our scheduling enhances performance by maximizing both intra-parallelism and inter-pipelining of attention head.
\begin{table}[t]
\caption{\revision{Simulation parameters for IANUS.}}
\label{Table: Simul param}
\begin{adjustbox}{width=1\linewidth,center}
\begin{tabular}{|c|c|c|}
\specialrule{1.25pt}{0pt}{0pt}
\multirow{3}{*}{NPU}    & Composition                                                         & 4 cores, 8 PIM memory controllers                                                                                                                         \\ \cline{2-3} 
                        & Host interface                                                      & PCIe 5.0 $\times$16                                                                                                                                              \\ \cline{2-3} 
                        & Frequency                                                           & 700 MHz                                                                                                                                                    \\ \hline
\multirow{5}{*}{Core} & Matrix unit                                                         & 128x64   processing elements (PEs), 4 MACs per PE, 46 TFLOPS                                                                                                \\ \cline{2-3} 
                        & Vector unit                                                         & Sixteen 4-wide VLIW processors     \\ \cline{2-3} 
                        & Scheduler                                                           & \begin{tabular}[c]{@{}c@{}}4   command slots per issue queue of units, \\256 command slots in pending queue  \end{tabular}                                                                                    \\ \cline{2-3} 
                        & Scratch-pad                                                         & Activation   12 MB, Weight 4 MB                                                                                                                             \\ \hline
\multirow{5}{*}{PIM}    & \begin{tabular}[c]{@{}c@{}}Memory\\      configuration\end{tabular} & \begin{tabular}[c]{@{}c@{}}GDDR6   16 Gb/s; $\times$16 organization; 8 channels; 256 GB/s; \\    2 channels per chip, 16 banks per channel, row (page) size 2 KB\end{tabular} \\ \cline{2-3} 
                        & Timing parameters                                                   & \begin{tabular}[c]{@{}c@{}} $t_{CK}=0.5 ns$, $t_{CCD_{S}}=t_{CCD_{L}}=1 ns$, $t_{RAS}=21 ns$, \\ $t_{WR}=36 ns$, $t_{RP}=30 ns$, $t_{RCDRD}=36 ns$, $t_{RCDWR}=24 ns$\end{tabular}                                             \\ \cline{2-3} 
                        & Processing unit (PU)                                                & 1 GHz; 1 PU per bank; 32 GFLOPS per PU                                                                                                                      \\
                        \cline{2-3}
                        & Global buffer                                    & One 2 KB global buffer per channel                                                                                                                      \\
                        \specialrule{1.25pt}{0pt}{0pt}
\end{tabular}
\end{adjustbox}
\end{table}

\begin{figure*}[t] %%% t: top, b: bottom, h: here
\begin{center}
\includegraphics[page=10,clip, trim=0cm 9.21cm 0cm 0cm,width=1.0\linewidth]{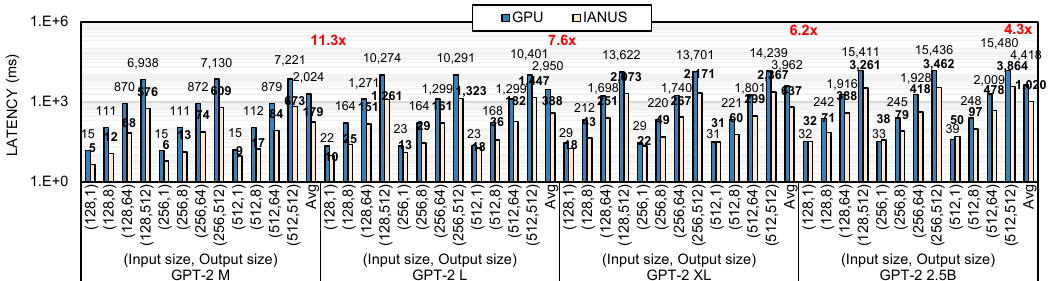}
\end{center}
\caption{Inference latency of various GPT-2 models on A100 GPU and IANUS.}
\label{fig:Latency GPT}
\end{figure*}

\begin{table}[t]
\caption{Specifications of A100 GPU, DFX~\cite{hong2022dfx}, and IANUS.}
\label{Table: Hardware specification}
\begin{adjustbox}{width=1\linewidth,center}
\begin{tabular}{|c|c|c|c|c|}
\specialrule{1.25pt}{0pt}{0pt}
                                                                          &                                                                     & A100~\cite{nvidia2021a100}           & \begin{tabular}[c]{@{}c@{}}DFX~\cite{hong2022dfx}\end{tabular} & IANUS                                                                                      \\
\hline
\hline
\multirowcell{2}[0pt][c]{Compute}         
& \begin{tabular}[c]{@{}c@{}}Frequency\end{tabular}           & 1155 MHz           & 200 MHz    & 700 MHz  \\ \cline{2-5} 
& \begin{tabular}[c]{@{}c@{}}Throughput\end{tabular}       & 255 TFLOPS            & 1.64 TFLOPS    & 184 TFLOPS \\ \hline
\begin{tabular}[c]{@{}c@{}}On-chip\\Memory\end{tabular}       & \begin{tabular}[c]{@{}c@{}}Capacity\\\end{tabular}             & RF, L1, L2: 84 MB & $\sim$40 MB                                                 & \begin{tabular}[c]{@{}c@{}}Activation Scratch-pad: 48 MB\\ Weight Scratch-pad: 16 MB\end{tabular} \\ \hline 
\multirowcell{4}[0pt][c]{\begin{tabular}[c]{@{}c@{}}Off-chip\\Memory\end{tabular}}       & Type                   & HBM2e          & HBM2     & GDDR6                              \\ \cline{2-5} 
                                                                        & \begin{tabular}[c]{@{}c@{}}Capacity\end{tabular}             & 80 GB             & 32 GB                                                       & 8 GB                                                                                          \\ \cline{2-5} 
                                                                          & \begin{tabular}[c]{@{}c@{}}Bandwidth\end{tabular}          & 2039 GB/s           & 1840 GB/s                                                     & 256 GB/s                                                                                        \\ \cline{2-5} 
                                                                          & \begin{tabular}[c]{@{}c@{}}Internal BW\end{tabular} & N/A              & N/A                                                        & 4096 GB/s    \\ 
\specialrule{1.25pt}{0pt}{0pt}
\end{tabular}
\end{adjustbox}
\end{table}

\section{Evaluations}

\subsection{Methodology}~\label{Sec:Evaluation Methodology}
\revision{To evaluate the performance of IANUS, we developed a cycle-accurate in-house simulator to model IANUS.  The simulator integrates an NPU simulator based on a commercial NPU~\cite{ahn2019aix,hwang2019aix,x330} as well as a PIM simulator modeled after the real PIM chip, AiM~\cite{lee20221ynm,kwon20221ynm}. Both the NPU and the PIM simulator are validated against their respective real hardware counterparts within a 5\% error margin.
An overview of the key simulation parameters is summarized in Table~\ref{Table: Simul param}.
In addition, we modeled the new components added to enable IANUS, including the PIM control unit (PCU), and modified the memory controller to support both PIM commands and normal memory commands.}
To avoid latency overhead from the PCU, we designed its operations to be pipelined with PIM computations.
Our simulator also provides statistics on energy consumption.
It measures the dynamic energy consumed by cores in NPU, PIM operations, and standard DRAM operations.
Based on prior analysis~\cite{kwon20221ynm}, we assume that the power consumption of PIM computing operations is 3$\times$ of that for DRAM read operations.

We compare the performance of IANUS against a GPU, state-of-the-art prior work (DFX~\cite{hong2022dfx}), as well as the NPU without PIM memory.
For the GPU, we use an NVIDIA A100-SXM GPU \cite{nvidia2021a100} with Pytorch 2.0 and CUDA Toolkit 11.8 and GPU-optimized source codes from Huggingface~\cite{wolf2019huggingface} and Megatron-LM~\cite{shoeybi2019megatron}.
The latency of models is measured using the \textit{torch.cuda.Event} API.
DFX~\cite{hong2022dfx} is a multi-FPGA appliance specifically designed to accelerate the \textit{generation} stage of GPT models.
We assume a DFX with 4 FPGAs that can support GPT-2 XL model.
We also compare IANUS with a commercial NPU~\cite{ahn2019aix,hwang2019aix,x330} (the same NPU used in IANUS) without PIM, but with standard GDDR6 memory (\texttt{NPU-MEM}). 
It shares identical specifications with IANUS in Table \ref{Table: Hardware specification} except for the internal memory bandwidth
and features a peak throughput of 184 TFLOPS.
IANUS is identical to NPU-MEM, except that standard GDDR6 memory is replaced with PIM based on AiM~\cite{lee20221ynm, kwon20221ynm}.
Each PIM chip achieves a peak throughput of 1 TFLOPS with 32 processing units utilizing 1024 GB/s internal memory bandwidth.
The specifications of each architecture are summarized in Table \ref{Table: Hardware specification}.

\begin{table}[t]
\centering
\caption{Network configuration details.}
\label{Table:Network Configuration}
\begin{adjustbox}{width=1.0\linewidth,center}
\begin{tabular}{|c|c|cccccc|}
\specialrule{1.25pt}{0pt}{0pt}
                      & Name & \begin{tabular}[c]{@{}c@{}}Embedding\\ dimension\end{tabular} & \begin{tabular}[c]{@{}c@{}}Head\\ dimension\end{tabular} & \# Heads & \# Blocks & \# Params & Workload                                                                       \\ \hline
\multirow{4}{*}{BERT} & B    & 768                                                           & 64                                                       & 12       & 12        & 110M      & \multirow{4}{*}{\begin{tabular}[c]{@{}c@{}}Question-\\ answering\\ (QA)\end{tabular}} \\ \cline{2-2}
                      & L    & 1024                                                          & 64                                                       & 16       & 24        & 340M      &                                                                                \\ \cline{2-2}
                      & 1.3B & 2048                                                          & 64                                                       & 32       & 24        & 1.3B      &                                                                                \\ \cline{2-2}
                      & 3.9B & 2560                                                          & 64                                                       & 40       & 48        & 3.9B      &                                                                                \\ \cline{1-8}
\multirow{4}{*}{GPT}  & M    & 1024                                                          & 64                                                       & 16       & 24        & 345M      & \multirow{4}{*}{\begin{tabular}[c]{@{}c@{}}Language\\ modeling\\ (LM)\end{tabular}}   \\ \cline{2-2}
                      & L    & 1280                                                          & 64                                                       & 20       & 36        & 762M      &                                                                                \\ \cline{2-2}
                      & XL   & 1536                                                          & 64                                                       & 24       & 48        & 1.5B      &                                                                                \\ \cline{2-2}
                      & 2.5B & 1920                                                          & 96                                                       & 20       & 54        & 2.5B      &                                                                                \\ \specialrule{1.25pt}{0pt}{0pt}
\end{tabular}
\end{adjustbox}
\end{table}

We evaluate two notable transformer-based LLMs, BERT \cite{devlin2018bert} and GPT \cite{radford2019language} with the BF16 \cite{wang2019bfloat16} data type, which maintains the accuracy of the full-precision model.
The configurations and tasks of each model are presented in Table \ref{Table:Network Configuration}.
We exploit a GPT-2 XL model with its attention heads reduced from 25 to 24, whose accuracy was validated in \cite{hong2022dfx}, to optimize parallelism.
We assess the end-to-end performance of models with input sizes of 128, 256, and 512 tokens.
For the GPT-2, we use output sizes of 1, 8, 64, and 512 tokens.
These sizes represent the typical user request ranges for NLP services in datacenters \cite{openai}.
Due to the time overhead associated with gathering inputs from multiple users, current datacenters prefer running the model with non-batched input \cite{fowers2018configurable,hong2022dfx}; therefore, we evaluate our work using a batch size of 1.

\subsection{Performance Results}~\label{sec: perf results}
\textbf{End-to-end Inference Latency:}
Figure \ref{fig:Latency GPT} presents the end-to-end latency of GPT-2 models on the GPU and IANUS.
The result shows that IANUS achieves a 4.3$\times$ speedup compared to the GPU for the 2.5B model, on average.
For the workload with significantly more output tokens than input tokens, i.e., (128,512), IANUS demonstrates 12.0$\times$, 8.1$\times$, and 6.6$\times$ lower latency than the GPU for the GPT-2 M, L, and XL models, respectively.
These substantial speedups are obtained from the high utilization of PIM chips' internal bandwidth of 4096 GB/s for matrix-vector multiplication in the \textit{generation} stage.
On average, IANUS takes about 5.7 ms per token for \textit{generation} stages of the GPT-2 2.5B model with configuration (128,64), while the GPU takes about 29.9 ms.

\begin{figure}[t] %%% t: top, b: bottom, h: here
\begin{center}
\includegraphics[page=13,clip, trim=0cm 11.4cm 7.9cm 0cm,width=1.0\linewidth]{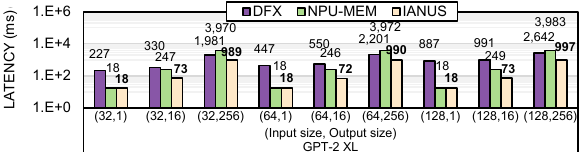}
\end{center}
\caption{Inference latency of GPT-2 XL on DFX~\cite{hong2022dfx}, NPU-MEM, and IANUS.}
\label{fig:Latency Compare}
\end{figure}

In Figure \ref{fig:Latency Compare}, we conduct a comparison of the GPT-2 XL's latency among IANUS, NPU-MEM, and DFX with four FPGAs \cite{hong2022dfx}, which provides state-of-the-art performance for GPT-2.
Input and output token sizes for the comparison are obtained from \cite{hong2022dfx}.
IANUS achieves a 49.3$\times$ speedup compared to DFX for the (128,1) configuration.
IANUS and NPU-MEM present similar performance for this configuration, as the PIM in IANUS operates as a standard GDDR6 except for the LM head.
For the \textit{generation} stage, DFX achieves 6.9 ms to generate one token for the (64,256) configuration, while IANUS generates a token in 3.8 ms for the same configuration, achieving a speedup of 1.8$\times$.
Without the benefits of PIM, NPU-MEM takes 15.5 ms.
To this end, IANUS achieves an average speedup of 3.2$\times$ compared to DFX, while NPU-MEM results in 24\% slowdown.

\begin{figure}[t] %%% t: top, b: bottom, h: here
\begin{center}
\includegraphics[page=20,clip, trim=0cm 11cm 7.9cm 0cm,width=1.0\linewidth]{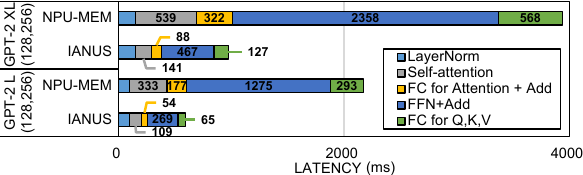}
\end{center}
\caption{Latency breakdown of GPT-2 XL and L's \textit{generation} stages for NPU-MEM and IANUS.}
\label{fig:NPU-MEM vs IANUS}
\end{figure}

\textbf{Latency Breakdown:}
To investigate the impact of using PIM, we measure the latency of operations in the decoder for NPU-MEM and IANUS in the \textit{generation} stages of GPT-2 L and XL.
As residual additions are executed with adjacent FC and FFN using a pipelining scheme, we collectively measure their latency.
As shown in Figure \ref{fig:NPU-MEM vs IANUS}, IANUS reduces the execution time of two FCs in multi-head attention from 890 ms to 215 ms for the GPT-2 XL model, achieving a speedup of 4.1$\times$. 
Since the FFN has a four times larger weight size compared to these two FCs, it achieves a higher speedup of 5.1 $\times$.
IANUS also achieves a speedup of 4.3 $\times$ for self-attention without offloading any operation in self-attention.
This speedup is obtained from prefetching previously generated keys and values instead of the weight for generating $Q, K,$ and $V$ by offloading FC for $Q, K,$ and $V$ generation to PIM.
Overall, IANUS achieves speedups of 4.0$\times$ and 3.6$\times$ for GPT-2 XL and L models, respectively, compared to NPU-MEM.

\begin{figure}[t] %%% t: top, b: bottom, h: here
\begin{center}
\includegraphics[page=17,clip, trim=0cm 11.4cm 7.9cm 0cm,width=1.0\linewidth]{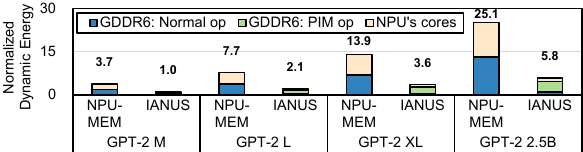}
\end{center}
\caption{Dynamic energy of NPU-MEM and IANUS, normalized to IANUS with GPT-2 M.}
\label{fig:Energy}
\end{figure}

\textbf{Energy Efficiency:}
Figure \ref{fig:Energy} presents dynamic energy consumption of IANUS and NPU-MEM for GPT-2 models where input and output token sizes are set to 256 and 512, respectively.
The energy values are normalized to the dynamic energy consumed by IANUS with GPT-2 M.
By offloading FC layers of the \textit{generation} stage to PIM, IANUS achieves 10.5-13.4$\times$ reduction in energy consumption for normal memory operations across all models.
The energy consumption for computation of cores in NPU is also decreased by a factor of 6.3-10.2$\times$.
The reduction in energy consumption for cores' computation and normal memory operations tends to increase as the model size expands.
Meanwhile, the energy is consumed by PIM operations in IANUS.
As a result, IANUS obtains 3.7$\times$, 3.6$\times$, 3.9$\times$, and 4.4$\times$ improvement in energy-efficiency compared to NPU-MEM for GPT-2 M, L, XL, and 2.5B, respectively.
Despite its larger model size, GPT-2 L results in a smaller energy efficiency improvement compared to GPT-2 M due to its embedding dimension size of 1280, which results in twice the number of row activations, compared to GPT-2 M's size of 1024 for PIM computation.
Note that energy efficiency in a real system can be further improved if static energy consumption is also considered~\footnote{Static energy consumption was not incorporated in the analysis because of the challenge in providing fair comparisons.}.

\begin{figure}[t] %%% t: top, b: bottom, h: here
\begin{center}
\includegraphics[page=24,clip, trim=0cm 12.1cm 7.9cm 0cm,width=1.0\linewidth]{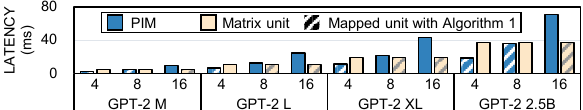}
\end{center}
\caption{Performance evaluation of the adaptive mapping algorithm for FC across different GPT-2 models as the number of input tokens are varied from 4, 8, to 16.}
\label{fig:verification workload}
\end{figure}

\textbf{Adaptive Mapping Algorithm for FC:}
To evaluate the benefits of Algorithm \ref{Algo: workload}, we evaluate the performance of GPT models when FC is mapped to PIM and matrix unit on various input token sizes and compared it to the result of Algorithm \ref{Algo: workload}.
As illustrated in Figure \ref{fig:verification workload}, Algorithm \ref{Algo: workload} \revision{is effective with small input sizes and} chooses the appropriate computation unit on various model sizes and input token sizes.
When executing FC layers in PIM, execution time is proportional to the input token size as PIM sequentially repeats matrix-vector multiplication as much as the input token size.
On the other hand, the matrix unit
shows similar performance across 4, 8, and 16 input tokens because of the capability of processing 128 tokens in parallel.
Therefore, the matrix unit achieves better performance for large input token sizes.
Another factor for workload mapping is the embedding size of the model.
As the global buffer and row size of PIM is 2KB (= 1024 BF16), models with embedding sizes that are multiples of 1024 can fully utilize PIM.
As a result, PIM shows higher performance than the matrix unit at an input size of 8 for GPT-2 M (embedding size of 1024) and GPT-2 2.5B (1920, nearly 2$\times$1024).
With Algorithm \ref{Algo: workload}, we achieve an average speedup of 1.4$\times$ and 1.2$\times$ when compared to mapping FC to PIM and the matrix unit, respectively.

\begin{figure}[t] %%% t: top, b: bottom, h: here
\begin{center}
\includegraphics[page=27,clip, trim=0cm 11.5cm 7.9cm 0.01cm,width=1.0\linewidth]{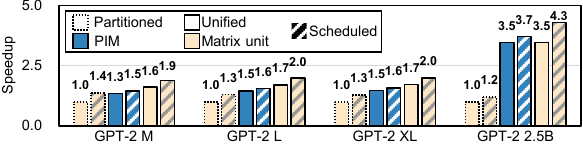}
\end{center}
\caption{Performance comparisons between unified and partitioned memory systems and the impact of unified memory-aware scheduling. Dashes in the bar border indicate the system type. Colors represent mapped units of $QK^T$ and $SV$. The pattern indicates the application of scheduling.}
\label{fig: Scheduling}
\end{figure}

\phantomsection~\label{sec: unified vs partitioned}
\textbf{Unified vs. Partitioned Memory System:}
In Figure~\ref{fig: Scheduling}, we evaluate the performance benefits of unified memory system (IANUS) compared to partitioned memory organization. Both configurations have the same total memory capacity of 8 GB -- with the unified system having 8 GB for both the PIM and the NPU, while for the partitioned configuration, 4 GB is dedicated for the NPU's main memory and 4 GB is dedicated for PIM. While the memory capacity is the same, the unified memory has the benefit of additional compute provided through the extra amount of PIM memory available.

We evaluate both systems using GPT-2 models with a (256,512) configuration.
\revision{
In the partitioned memory, all FC parameters shared between PIMs and NPU are duplicated across the both memory to avoid performance overhead caused by data movement between standard DRAMs and PIMs.
However, for the 2.5B model, the entire FC parameters cannot be duplicated across both the PIM and the DRAM because of the limited memory capacity.  
Thus, to minimize transfer overhead between the two types of memories, the NPU's matrix unit is mainly responsible for the FC operations on the non-duplicated parameters.
}
For a fair comparison, we implement scheduling for the partitioned memory system that maximizes the benefits from parallel executions of NPU and PIM by mapping the $QK^T$ and $SV$ to the matrix unit.
As shown in Figure \ref{fig: Scheduling}, the concurrent execution of NPU's DRAM accesses and PIM computations results in an average 1.3$\times$ speedup in the partitioned system.

\revision{For GPT-2 M, L, and XL models,}
IANUS--the unified memory system--(the rightmost bar for each model) outperforms the scheduled partitioned memory system by 1.4-1.6$\times$ speedup (\revision{Figure \ref{fig: Scheduling}).
These speedups result from the doubled PIM throughput that is available in the unified memory configuration. 
For the GPT-2 2.5B model, IANUS shows a larger performance improvement due to the performance overhead in the partitioned system, stemming from the data movement of non-duplicated parameters from the PIM to the NPU.
Similar to performance trends of other models, while not shown, IANUS achieves approximately 1.5$\times$ speedup in GPT-2 2.5B compared to the partitioned system if sufficient memory capacity is provided such that all FC parameters can be stored in each memory type.}

\textbf{Unified Memory-Aware Scheduling for Multi-Head Attention:}
Figure \ref{fig: Scheduling} demonstrates the performance enhancement through mapping of $QK^T$ and $SV$ operations and corresponding scheduling for multi-head attention in IANUS (the unified memory system).
As in the figure, scheduling for the mapping of $QK^T$ and $SV$ to PIM results in an average performance boost of 7\% across all models compared to naïve scheduling.
When $QK^T$ and $SV$ operations are mapped to the matrix unit, a reduction in computation time for these operations leads to superior performance than the case of scheduling with PIM mapping for all models except GPT-2 2.5B.
For the GPT-2 2.5B model, which has a larger head dimension size of 96 than other models, the loading time for the previously generated keys and values increases.
This loading time is not required when $QK^T$ and $SV$ are mapped to PIM, thus reducing the benefits gained through matrix unit mapping.
However, through effective scheduling, we attain a performance improvement of 24\% for GPT-2 2.5B.
Consequently, unified memory-aware scheduling yields an average performance improvement of 34\%.

\begin{figure}[t] %%% t: top, b: bottom, h: here
\begin{center}
\includegraphics[page=4,clip, trim=0cm 10.7cm 7.9cm 0cm,width=1.0\linewidth]{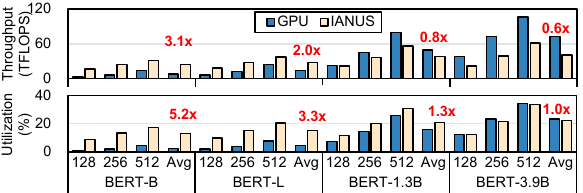}
\end{center}
\caption{Throughput and compute utilization of the BERT models on A100 GPU and IANUS.}
\label{fig:Latency BERT}
\end{figure}

\textbf{Throughput and Compute Utilization:}
Figure \ref{fig:Latency BERT} presents the throughput and utilization of the IANUS and the GPU for BERT models.
In IANUS, only the matrix unit and vector unit of the NPU are utilized for computation, excluding PIM, as BERT models do not include matrix-vector multiplication.
By managing complex data manipulation in self-attention through on-chip data movement, IANUS attains 3.1$\times$ and 2.0$\times$ higher average throughput for BERT-B and L, respectively, despite having 1.4$\times$ lower peak FLOPS than the GPU.

As the FLOPs increase with model size, IANUS's throughput becomes less than the GPU due to its limited peak FLOPS.
However, IANUS achieves 5.2$\times$, 3.3$\times$, 1.3$\times$, and 1.0$\times$ higher average utilization for BERT-B, L, 1.3B, and 3.9B compared to the GPU.
This enhanced utilization is attributed to the efficient execution of vector operations with the vector unit in addition to the benefits gained from self-attention.

\begin{figure}[t] %%% t: top, b: bottom, h: here
\begin{center}
\includegraphics[page=7,clip, trim=0cm 12.2cm 7.9cm 0cm,width=1.0\linewidth]{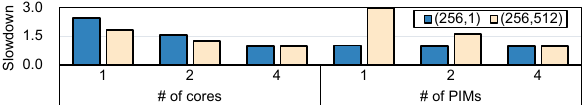}
\end{center}
\caption{Sensitivity studies for \textit{summarization}-only (256,1) and \textit{generation}-dominant cases (256,512) as the numbers of cores and PIM chips are varied. Results are normalized to 4 cores and 4 PIMs.}
\label{fig:sensitivity}
\end{figure}

\textbf{Sensitivity Study of Design Parameters:}
We conduct sensitivity studies on the number of cores in NPU and PIM chips.
To show the sensitivities of NPU and PIM computation capabilities, we keep memory bandwidth the same as the baseline while varying the number of cores and PIM chips.
We present \textit{summarization}-only (256,1) and \textit{generation}-dominant (256,512) cases for comprehensive analysis with GPT-2 L to isolate the impacts from reduced on-chip memory or PIM capacity.
As shown in Figure~\ref{fig:sensitivity}, the fewer cores result in slowdowns for both cases due to the decreased intra-layer and attention-head parallelism, and \textit{summarization}-only case suffers more as NPU executes all but one computation (LM head). 
On the other hand, PIM's computation capability significantly affects the \textit{generation}-dominant configuration, where a significant fraction of FC operations are executed on PIMs.

\begin{figure}[t] %%% t: top, b: bottom, h: here
\begin{center}
\includegraphics[page=29,clip, trim=0cm 10.7cm 7.9cm 0cm,width=1.0\linewidth]{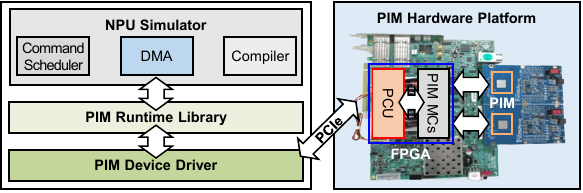}
\end{center}
\caption{System prototype of IANUS (PCU: PIM Control Unit).}
\label{fig:prototype}
\end{figure}

\subsection{IANUS System Prototyping}
We develop a system prototype of IANUS to validate feasibility as shown in Figure \ref{fig:prototype}. 
Our prototype is based on commodity Xilinx FPGA board (VCU118) \cite{vcu118} to assess the feasibility of IANUS with real PIM chips, GDDR6-AiM \cite{lee20221ynm,kwon20221ynm}.
Specifically, we use the AiM-embedded FPGA Mezzanine Card (FMC) and connect it to the FPGA via the FMC connector~\cite{kwon2022system}.
% , through FMC connector~\cite{kwon2022system}. 
As shown, the PIM control unit (PCU) and PIM memory controllers (PIM MCs) are implemented on FPGA whereas we leverage our NPU simulator as the NPU of IANUS since its RTL design was too big to fit in a single FPGA. 
When macro PIM commands are ready to be executed in the NPU, they are dispatched to PCU through the PCIe interface by the PIM runtime library and device driver. These macro commands are then converted into corresponding micro PIM commands and are transferred to PIM through PCU and PIM MCs. DMA commands from the NPU simulator are also similarly transferred to PIM.
To validate the functionality of a system prototype for IANUS, we evaluate the accuracy of our system using pretrained models of GPT-2 \cite{radford2019language} on the WikiText-2 dataset.
Our system prototype achieves perplexity scores of 30.92 and 22.60, 19.39, and 17.48 for GPT-2 Base (117M), M, L, and XL, respectively, achieving similar perplexity scores as the full-precision models.

\section{Discussion} 
\begin{table}[t]
\centering
\caption{\revision{Network configurations of larger LLMs.}}
\label{Table:config Large LLM}
\begin{adjustbox}{width=1.0\linewidth,center}
\begin{tabular}{|c|cccccc|}
\specialrule{1.25pt}{0pt}{0pt}
                      & \# Params & \begin{tabular}[c]{@{}c@{}}Embedding\\ dimension\end{tabular} & \begin{tabular}[c]{@{}c@{}}Head\\ dimension\end{tabular} & \# Heads & \# Blocks & Workload                                                                       \\ \hline
\multirow{3}{*}{GPT}     & 6.7B & 4096                                                          & 128                                                       & 32       & 32         & \multirow{3}{*}{\begin{tabular}[c]{@{}c@{}}Language\\ modeling\\ (LM)\end{tabular}} \\
                         & 13B & 5120                                                          & 128                                                       & 40       & 40              &                                                                                \\ 
                         & 30B & 7168                                                          & 128                                                       & 56       & 48    &                                         \\ \specialrule{1.25pt}{0pt}{0pt}
\end{tabular}
\end{adjustbox}
\end{table}

\begin{figure}[t] %%% t: top, b: bottom, h: here
\begin{center}
\includegraphics[page=31,clip, trim=0cm 9.7cm 7.9cm 0cm,width=1.0\linewidth]{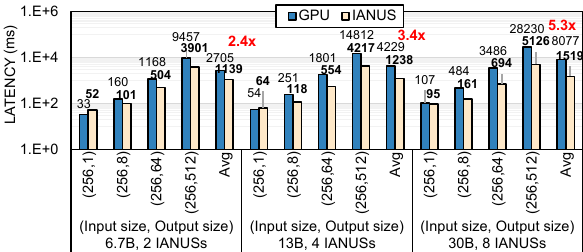}
\end{center}
\caption{\revision{Inference performance scalability for larger LLMs with multiple IANUS devices.  The results are compared to a single A100 GPU.}}
\label{fig:Larger LLM}
\end{figure}

\subsection{Scalability Analysis}~\label{Sec: Scalability}
Given the limited memory capacity of IANUS compared to modern GPUs, the memory capacity of IANUS needs to be scaled to run larger LLMs.
The memory (PIM) capacity of IANUS can be expanded in two ways: 1) increase the amount of PIM per NPU, or 2) scale the number of IANUS devices. The first approach can be achieved by adding more PIM controllers or employing a clamshell configuration of GDDR6 devices \cite{gddr6}.
However, this approach requires modifications to the IANUS architecture.
In this work, we leverage the second approach to analyze the scalability of IANUS. 

\revision{
The larger LLMs used in the scalability analysis are summarized in Table~\ref{Table:config Large LLM}. 
For each model, the number of IANUS devices is selected to provide sufficient memory capacity to support the model -- i.e.,  two, four, and eight IANUS devices are used to support the GPT 6.7B, 13B, and 30B models, respectively.
The multiple IANUS devices are assumed to be interconnected through PCIe 5.0 $\times$16 host interface. 
To maximize parallelism across IANUS devices, both intra-layer parallelism and attention head parallelism are exploited among devices.
}

\revision{As shown in Figure \ref{fig:Larger LLM}, multiple IANUS devices provide average speedups of 2.4$\times$, 3.4$\times$, and 5.3$\times$ across respective models, compared to a single A100 GPU, which has sufficient memory capacity for the larger LLMs.
Multiple IANUS devices not only provide additional memory capacity but also increase effective memory bandwidth with extra PIM capability.
For larger LLMs, the key system component that impacts overall performance is the memory bandwidth.
This is because the proportion of FC layers in LLMs, which are bottlenecked by memory bandwidth, increases as the size of the LLMs grows.
Leveraging the PIM’s internal memory bandwidth, the effective memory bandwidth of IANUS reaches approximately 2.4 TB/s, 9-10$\times$ higher than external memory bandwidth of GDDR6 memories.
Thus, with two IANUS devices, the total effective bandwidth is 4.8 TB/s, which is approximately 2.4$\times$ higher than the A100 memory bandwidth (2039 GB/s).
This difference in memory bandwidth nearly matches the observed performance benefits of two IANUS devices over the A100 GPU.
However, scaling the number of IANUS devices comes at the cost of communication overhead between IANUS devices compared to a single GPU. As a result, the performance benefits with four and eight IANUS devices do not match the theoretical memory bandwidth difference; however, there is still significant speedup compared to a single A100 GPU. 
}

\begin{figure}[t] %%% t: top, b: bottom, h: here
\begin{center}
\includegraphics[page=38,clip, trim=0cm 11.9cm 7.9cm 0cm,width=1.0\linewidth]{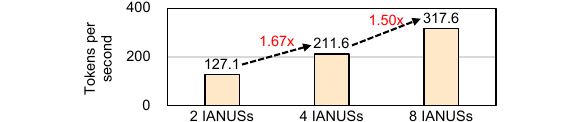}
\end{center}
\caption{\revision{Strong scaling of IANUS on the GPT 6.7B model.}}
\label{fig:scalability}
\end{figure}

\revision{Strong scaling of IANUS is shown in Figure~\ref{fig:scalability},  using the 6.7B model with a 256:64 token configuration.
As described earlier, the additional IANUS devices provide higher effective memory bandwidth and result in performance gain -- 2.5$\times$ performance improvement when the number of IANUS is increased by 4$\times$.
While the performance of IANUS improves with extra devices, 
linear speedup is not obtained because of the communication overhead between multiple devices.
A multi-IANUS device system presents new opportunities for optimizing communication across the devices but we leave such exploration as part of future work.
}

\subsection{Cost Analysis} \label{Sec: Cost}

\rindex{R2b}\revision{
Providing a fair cost comparison between two different system architectures (e.g., HBM memory with interposer vs GDDR6-based PIM memory) is a challenge since many factors impact cost.  However, prior work has shown how thermal design power (TDP) can approximate total cost of ownership (TCO) in datacenters~\cite{ten_lessons_tpu}. Therefore, we use TDP for the cost comparison.  The TDP of A100 GPU is estimated at 400 W~\cite{nvidia2021a100} while the TDP of IANUS is conservatively assumed to be 120 W, based on estimates from the NPU~\cite{ahn2019aix,hwang2019aix,x330} and PIM~\cite{lee20221ynm,kwon20221ynm} components. 
Using the performance/TDP metric for the cost-efficiency evaluation, 
configurations of two, four, and eight IANUS devices yield improvements in cost-efficiency of 3.9$\times$, 2.7$\times$, and 2.1$\times$ over the single A100 GPU for the 6.7B, 13B, and 30B models, respectively. 
For each comparison, the performances of IANUS devices and the GPU for the performance/TDP metric are measured with a 256:64 input-to-output token ratio.
While the cost-efficiency benefits of IANUS devices are evident, they diminish as the number of IANUS devices increases.
The cost efficiency of IANUS can potentially be enhanced by leveraging PIM chips with higher memory capacity and/or more PIM chips connected to a single NPU. 
}

\section{Related Works}
\textbf{Domain-specific Accelerators:}
Various hardware accelerators have been proposed to accelerate transformer models.
TurboTransformer \cite{fang2021turbotransformers} executes BERT variants effectively by operation fusion and pipelining.
Unfortunately, this approach suffers from severe under-utilization in text generation workloads.
Several accelerators \cite{ham20203,wang2021spatten,ham2021elsa,lu2021sanger,SparseMICRO22} focus on multi-head-attention mechanisms only, requiring additional hardware to handle other operations such as layer normalization. Meanwhile, IANUS utilizes integrated both NPU and PIM to accelerate end-to-end inference of LLMs.

\textbf{PIM Accelerators}: Utilizing large in-memory bandwidth, PIM architectures reduce massive data movement between DRAMs and a host \cite{kwon202125,li2017drisa,he2020newton,lee20221ynm,kim2022aquabolt,lee2021hardware}. McDRAM \cite{shin2018mcdram} presents near-memory structures and a horizontal arrangement of data within memory banks.
TransPIM~\cite{zhou2022transpim} introduces the first memory-based accelerator for end-to-end inference of transformers.
However, it only achieves an average throughput of 734 GOPS
due to area and power constraints.
\drindex{R2e}\revision{Chopim~\cite{Chopim} employs a unified memory system between the near-data accelerator and the host. Unlike Chopim, which focuses on homogeneous kernels,
IANUS targets heterogeneous kernels that require scheduling to optimize performance.}

\section{Conclusion}
We propose IANUS, an integrated accelerator based on NPU-PIM unified memory system, that fully exploits the benefits of NPU and PIM to accelerate end-to-end inference in transformer-based LLMs.
To overcome the challenges posed by a unified memory system with PIM, we propose \emph{PIM Access Scheduling} that schedules PIM operations and normal memory accesses through the workload mapping and scheduling.
IANUS results in 6.2$\times$ and 3.2$\times$ speedup in comparison to GPU and DFX-based solutions for the GPT-2 model, highlighting the potential of such hybrid architectures.
To demonstrate the feasibility of IANUS, we constructed an FPGA prototype system based on a commercial NPU and real PIM chips.

\begin{acks}
We thank the shepherd and all reviewers for their valuable comments. 
This work was supported in part by the IITP grant funded by the MSIT (No.RS 2023-00228255, PIM-NPU Based Processing System Software Developments for Hyper-scale Artificial Neural Network Processing) and in part by the IITP grant funded by the MSIT (No. 2021-0-00106, AI accelerator-optimized neural network automatic generation technology and open service platform development).
\end{acks}

\bibliographystyle{plain}
\bibliography{references}

\begin{thebibliography}{10}

\bibitem{ahn2019aix}
Minwook Ahn, Seok~Joong Hwang, Wonsub Kim, Seungrok Jung, Yeonbok Lee, Mookyoung Chung, Woohyung Lim, and Youngjoon Kim.
\newblock Aix: A high performance and energy efficient inference accelerator on fpga for a dnn-based commercial speech recognition.
\newblock In {\em 2019 Design, Automation \& Test in Europe Conference \& Exhibition (DATE)}, pages 1495--1500. IEEE, 2019.

\bibitem{albericio2016cnvlutin}
Jorge Albericio, Patrick Judd, Tayler Hetherington, Tor Aamodt, Natalie~Enright Jerger, and Andreas Moshovos.
\newblock Cnvlutin: Ineffectual-neuron-free deep neural network computing.
\newblock {\em ACM SIGARCH Computer Architecture News}, 44(3):1--13, 2016.

\bibitem{ba2016layer}
Jimmy~Lei Ba, Jamie~Ryan Kiros, and Geoffrey~E Hinton.
\newblock Layer normalization.
\newblock {\em arXiv preprint arXiv:1607.06450}, 2016.

\bibitem{bridle1989training}
John Bridle.
\newblock Training stochastic model recognition algorithms as networks can lead to maximum mutual information estimation of parameters.
\newblock {\em Advances in neural information processing systems}, 2, 1989.

\bibitem{chen2014diannao}
Tianshi Chen, Zidong Du, Ninghui Sun, Jia Wang, Chengyong Wu, Yunji Chen, and Olivier Temam.
\newblock Diannao: A small-footprint high-throughput accelerator for ubiquitous machine-learning.
\newblock {\em ACM SIGARCH Computer Architecture News}, 42(1):269--284, 2014.

\bibitem{chen2016eyeriss}
Yu-Hsin Chen, Tushar Krishna, Joel~S Emer, and Vivienne Sze.
\newblock Eyeriss: An energy-efficient reconfigurable accelerator for deep convolutional neural networks.
\newblock {\em IEEE journal of solid-state circuits}, 52(1):127--138, 2016.

\bibitem{Chopim}
Benjamin~Y. Cho, Yongkee Kwon, Sangkug Lym, and Mattan Erez.
\newblock Near data acceleration with concurrent host access.
\newblock In {\em 2020 ACM/IEEE 47th Annual International Symposium on Computer Architecture (ISCA)}, pages 818--831, 2020.

\bibitem{devaux2019true}
Fabrice Devaux.
\newblock The true processing in memory accelerator.
\newblock In {\em 2019 IEEE Hot Chips 31 Symposium (HCS)}, pages 1--24. IEEE Computer Society, 2019.

\bibitem{devlin2018bert}
Jacob Devlin, Ming-Wei Chang, Kenton Lee, and Kristina Toutanova.
\newblock Bert: Pre-training of deep bidirectional transformers for language understanding.
\newblock {\em arXiv preprint arXiv:1810.04805}, 2018.

\bibitem{fang2021turbotransformers}
Jiarui Fang, Yang Yu, Chengduo Zhao, and Jie Zhou.
\newblock Turbotransformers: an efficient gpu serving system for transformer models.
\newblock In {\em Proceedings of the 26th ACM SIGPLAN Symposium on Principles and Practice of Parallel Programming}, pages 389--402, 2021.

\bibitem{fisher1983very}
Joseph~A Fisher.
\newblock Very long instruction word architectures and the eli-512.
\newblock In {\em Proceedings of the 10th annual international symposium on Computer architecture}, pages 140--150, 1983.

\bibitem{fowers2018configurable}
Jeremy Fowers, Kalin Ovtcharov, Michael Papamichael, Todd Massengill, Ming Liu, Daniel Lo, Shlomi Alkalay, Michael Haselman, Logan Adams, Mahdi Ghandi, Stephen Heil, Prerak Patel, Adam Sapek, Gabriel Weisz, Lisa Woods, Sitaram Lanka, Steven~K. Reinhardt, Adrian~M. Caulfield, Eric~S. Chung, and Doug Burger.
\newblock A configurable cloud-scale dnn processor for real-time ai.
\newblock In {\em 2018 ACM/IEEE 45th Annual International Symposium on Computer Architecture (ISCA)}, pages 1--14, 2018.

\bibitem{ham20203}
Tae~Jun Ham, Sung~Jun Jung, Seonghak Kim, Young~H. Oh, Yeonhong Park, Yoonho Song, Jung-Hun Park, Sanghee Lee, Kyoung Park, Jae~W. Lee, and Deog-Kyoon Jeong.
\newblock $a^3$: Accelerating attention mechanisms in neural networks with approximation.
\newblock In {\em 2020 IEEE International Symposium on High Performance Computer Architecture (HPCA)}, pages 328--341, 2020.

\bibitem{ham2021elsa}
Tae~Jun Ham, Yejin Lee, Seong~Hoon Seo, Soosung Kim, Hyunji Choi, Sung~Jun Jung, and Jae~W Lee.
\newblock Elsa: Hardware-software co-design for efficient, lightweight self-attention mechanism in neural networks.
\newblock In {\em 2021 ACM/IEEE 48th Annual International Symposium on Computer Architecture (ISCA)}, pages 692--705. IEEE, 2021.

\bibitem{he2016deep}
Kaiming He, Xiangyu Zhang, Shaoqing Ren, and Jian Sun.
\newblock Deep residual learning for image recognition.
\newblock In {\em Proceedings of the IEEE conference on computer vision and pattern recognition}, pages 770--778, 2016.

\bibitem{he2020newton}
Mingxuan He, Choungki Song, Ilkon Kim, Chunseok Jeong, Seho Kim, Il~Park, Mithuna Thottethodi, and TN~Vijaykumar.
\newblock Newton: A dram-maker’s accelerator-in-memory (aim) architecture for machine learning.
\newblock In {\em 2020 53rd Annual IEEE/ACM International Symposium on Microarchitecture (MICRO)}, pages 372--385. IEEE, 2020.

\bibitem{hegde2018ucnn}
Kartik Hegde, Jiyong Yu, Rohit Agrawal, Mengjia Yan, Michael Pellauer, and Christopher Fletcher.
\newblock Ucnn: Exploiting computational reuse in deep neural networks via weight repetition.
\newblock In {\em 2018 ACM/IEEE 45th Annual International Symposium on Computer Architecture (ISCA)}, pages 674--687. IEEE, 2018.

\bibitem{hendrycks2016gaussian}
Dan Hendrycks and Kevin Gimpel.
\newblock Gaussian error linear units (gelus).
\newblock {\em arXiv preprint arXiv:1606.08415}, 2016.

\bibitem{hong2022dfx}
Seongmin Hong, Seungjae Moon, Junsoo Kim, Sungjae Lee, Minsub Kim, Dongsoo Lee, and Joo-Young Kim.
\newblock Dfx: A low-latency multi-fpga appliance for accelerating transformer-based text generation.
\newblock In {\em 2022 55th IEEE/ACM International Symposium on Microarchitecture (MICRO)}, pages 616--630. IEEE, 2022.

\bibitem{hwang2019aix}
Seok~Joong Hwang, Jeongho Han, Minwook Ahn, Seungrok Jung, Wonsub Kim, Yongshik Moon, Sangjun Yang, Moo-Kyoung Chung, Jaehyeok Jang, Youngjae Jin, Yongsang Park, Namseob Lee, Daewoo Kim, Euiseok Kim, Choong~Hwan Choi, and Heeyul Lee.
\newblock Aix v2: Flexible high performance ai inference accelerator for datacenters.
\newblock In {\em 2019 IEEE Hot Chips 31 Symposium (HCS)}, 2019.

\bibitem{isca-tpuv4}
Norm Jouppi, George Kurian, Sheng Li, Peter Ma, Rahul Nagarajan, Lifeng Nai, Nishant Patil, Suvinay Subramanian, Andy Swing, Brian Towles, Clifford Young, Xiang Zhou, Zongwei Zhou, and David~A Patterson.
\newblock Tpu v4: An optically reconfigurable supercomputer for machine learning with hardware support for embeddings.
\newblock In {\em Proceedings of the 50th Annual International Symposium on Computer Architecture}, ISCA '23, New York, NY, USA, 2023. Association for Computing Machinery.

\bibitem{ten_lessons_tpu}
Norman~P. Jouppi, Doe Hyun~Yoon, Matthew Ashcraft, Mark Gottscho, Thomas~B. Jablin, George Kurian, James Laudon, Sheng Li, Peter Ma, Xiaoyu Ma, Thomas Norrie, Nishant Patil, Sushma Prasad, Cliff Young, Zongwei Zhou, and David Patterson.
\newblock Ten lessons from three generations shaped google’s tpuv4i : Industrial product.
\newblock In {\em 2021 ACM/IEEE 48th Annual International Symposium on Computer Architecture (ISCA)}, pages 1--14, 2021.

\bibitem{jouppi2017datacenter}
Norman~P. Jouppi, Cliff Young, Nishant Patil, David Patterson, Gaurav Agrawal, Raminder Bajwa, Sarah Bates, Suresh Bhatia, Nan Boden, Al~Borchers, Rick Boyle, Pierre-luc Cantin, Clifford Chao, Chris Clark, Jeremy Coriell, Mike Daley, Matt Dau, Jeffrey Dean, Ben Gelb, Tara~Vazir Ghaemmaghami, Rajendra Gottipati, William Gulland, Robert Hagmann, C.~Richard Ho, Doug Hogberg, John Hu, Robert Hundt, Dan Hurt, Julian Ibarz, Aaron Jaffey, Alek Jaworski, Alexander Kaplan, Harshit Khaitan, Daniel Killebrew, Andy Koch, Naveen Kumar, Steve Lacy, James Laudon, James Law, Diemthu Le, Chris Leary, Zhuyuan Liu, Kyle Lucke, Alan Lundin, Gordon MacKean, Adriana Maggiore, Maire Mahony, Kieran Miller, Rahul Nagarajan, Ravi Narayanaswami, Ray Ni, Kathy Nix, Thomas Norrie, Mark Omernick, Narayana Penukonda, Andy Phelps, Jonathan Ross, Matt Ross, Amir Salek, Emad Samadiani, Chris Severn, Gregory Sizikov, Matthew Snelham, Jed Souter, Dan Steinberg, Andy Swing, Mercedes Tan, Gregory Thorson, Bo~Tian, Horia Toma, Erick Tuttle, Vijay
  Vasudevan, Richard Walter, Walter Wang, Eric Wilcox, and Doe~Hyun Yoon.
\newblock In-datacenter performance analysis of a tensor processing unit.
\newblock In {\em Proceedings of the 44th Annual International Symposium on Computer Architecture}, ISCA '17, page 1–12, New York, NY, USA, 2017. Association for Computing Machinery.

\bibitem{kim2022aquabolt}
Jin~Hyun Kim, Shin-Haeng Kang, Sukhan Lee, Hyeonsu Kim, Yuhwan Ro, Seungwon Lee, David Wang, Jihyun Choi, Jinin So, YeonGon Cho, JoonHo Song, Jeonghyeon Cho, Kyomin Sohn, and Nam~Sung Kim.
\newblock Aquabolt-xl hbm2-pim, lpddr5-pim with in-memory processing, and axdimm with acceleration buffer.
\newblock {\em IEEE Micro}, 42(3):20--30, 2022.

\bibitem{kung1979systolic}
Hsiang~Tsung Kung and Charles~E Leiserson.
\newblock Systolic arrays (for vlsi).
\newblock In {\em Sparse Matrix Proceedings 1978}, volume~1, pages 256--282. Society for industrial and applied mathematics Philadelphia, PA, USA, 1979.

\bibitem{kwon20221ynm}
Daehan Kwon, Seongju Lee, Kyuyoung Kim, Sanghoon Oh, Joonhong Park, Gi-Moon Hong, Dongyoon Ka, Kyudong Hwang, Jeongje Park, Kyeongpil Kang, Jungyeon Kim, Junyeol Jeon, Nahsung Kim, Yongkee Kwon, Vladimir Kornijcuk, Woojae Shin, Jongsoon Won, Minkyu Lee, Hyunha Joo, Haerang Choi, Guhyun Kim, Byeongju An, Jaewook Lee, Donguc Ko, Younggun Jun, Ilwoong Kim, Choungki Song, Ilkon Kim, Chanwook Park, Seho Kim, Chunseok Jeong, Euicheol Lim, Dongkyun Kim, Jieun Jang, Il~Park, Junhyun Chun, and Joohwan Cho.
\newblock A 1ynm 1.25 v 8gb 16gb/s/pin gddr6-based accelerator-in-memory supporting 1tflops mac operation and various activation functions for deep learning application.
\newblock {\em IEEE Journal of Solid-State Circuits}, 58(1):291--302, 2022.

\bibitem{kwon2018maeri}
Hyoukjun Kwon, Ananda Samajdar, and Tushar Krishna.
\newblock Maeri: Enabling flexible dataflow mapping over dnn accelerators via reconfigurable interconnects.
\newblock {\em ACM SIGPLAN Notices}, 53(2):461--475, 2018.

\bibitem{kwon2022system}
Yongkee Kwon, Kornijcuk Vladimir, Nahsung Kim, Woojae Shin, Jongsoon Won, Minkyu Lee, Hyunha Joo, Haerang Choi, Guhyun Kim, Byeongju An, Jeongbin Kim, Jaewook Lee, Ilkon Kim, Jaehan Park, Chanwook Park, Yosub Song, Byeongsu Yang, Hyungdeok Lee, Seho Kim, Daehan Kwon, Seongju Lee, Kyuyoung Kim, Sanghoon Oh, Joonhong Park, Gimoon Hong, Dongyoon Ka, Kyudong Hwang, Jeongje Park, Kyeongpil Kang, Jungyeon Kim, Junyeol Jeon, Myeongjun Lee, Minyoung Shin, Minhwan Shin, Jaekyung Cha, Changson Jung, Kijoon Chang, Chunseok Jeong, Euicheol Lim, Il~Park, Junhyun Chun, and Sk~Hynix.
\newblock System architecture and software stack for gddr6-aim.
\newblock In {\em 2022 IEEE Hot Chips 34 Symposium (HCS)}, pages 1--25. IEEE, 2022.

\bibitem{kwon202125}
Young-Cheon Kwon, Suk~Han Lee, Jaehoon Lee, Sang-Hyuk Kwon, Je~Min Ryu, Jong-Pil Son, O~Seongil, Hak-Soo Yu, Haesuk Lee, Soo~Young Kim, Youngmin Cho, Jin~Guk Kim, Jongyoon Choi, Hyun-Sung Shin, Jin Kim, BengSeng Phuah, HyoungMin Kim, Myeong~Jun Song, Ahn Choi, Daeho Kim, SooYoung Kim, Eun-Bong Kim, David Wang, Shinhaeng Kang, Yuhwan Ro, Seungwoo Seo, JoonHo Song, Jaeyoun Youn, Kyomin Sohn, and Nam~Sung Kim.
\newblock 25.4 a 20nm 6gb function-in-memory dram, based on hbm2 with a 1.2 tflops programmable computing unit using bank-level parallelism, for machine learning applications.
\newblock In {\em 2021 IEEE International Solid-State Circuits Conference (ISSCC)}, volume~64, pages 350--352. IEEE, 2021.

\bibitem{lee20221ynm}
Seongju Lee, Kyuyoung Kim, Sanghoon Oh, Joonhong Park, Gimoon Hong, Dongyoon Ka, Kyudong Hwang, Jeongje Park, Kyeongpil Kang, Jungyeon Kim, Junyeol Jeon, Nahsung Kim, Yongkee Kwon, Kornijcuk Vladimir, Woojae Shin, Jongsoon Won, Minkyu Lee, Hyunha Joo, Haerang Choi, Jaewook Lee, Donguc Ko, Younggun Jun, Keewon Cho, Ilwoong Kim, Choungki Song, Chunseok Jeong, Daehan Kwon, Jieun Jang, Il~Park, Junhyun Chun, and Joohwan Cho.
\newblock A 1ynm 1.25 v 8gb, 16gb/s/pin gddr6-based accelerator-in-memory supporting 1tflops mac operation and various activation functions for deep-learning applications.
\newblock In {\em 2022 IEEE International Solid-State Circuits Conference (ISSCC)}, volume~65, pages 1--3. IEEE, 2022.

\bibitem{lee2021hardware}
Sukhan Lee, Shin-haeng Kang, Jaehoon Lee, Hyeonsu Kim, Eojin Lee, Seungwoo Seo, Hosang Yoon, Seungwon Lee, Kyounghwan Lim, Hyunsung Shin, Jinhyun Kim, O~Seongil, Anand Iyer, David Wang, Kyomin Sohn, and Nam~Sung Kim.
\newblock Hardware architecture and software stack for pim based on commercial dram technology : Industrial product.
\newblock In {\em 2021 ACM/IEEE 48th Annual International Symposium on Computer Architecture (ISCA)}, pages 43--56, 2021.

\bibitem{li2017drisa}
Shuangchen Li, Dimin Niu, Krishna~T Malladi, Hongzhong Zheng, Bob Brennan, and Yuan Xie.
\newblock Drisa: A dram-based reconfigurable in-situ accelerator.
\newblock In {\em Proceedings of the 50th Annual IEEE/ACM International Symposium on Microarchitecture}, pages 288--301, 2017.

\bibitem{liu2015pudiannao}
Daofu Liu, Tianshi Chen, Shaoli Liu, Jinhong Zhou, Shengyuan Zhou, Olivier Teman, Xiaobing Feng, Xuehai Zhou, and Yunji Chen.
\newblock Pudiannao: A polyvalent machine learning accelerator.
\newblock {\em ACM SIGARCH Computer Architecture News}, 43(1):369--381, 2015.

\bibitem{lu2021sanger}
Liqiang Lu, Yicheng Jin, Hangrui Bi, Zizhang Luo, Peng Li, Tao Wang, and Yun Liang.
\newblock Sanger: A co-design framework for enabling sparse attention using reconfigurable architecture.
\newblock In {\em MICRO-54: 54th Annual IEEE/ACM International Symposium on Microarchitecture}, pages 977--991, 2021.

\bibitem{gddr6}
Micron.
\newblock Gddr6 datasheet.
\newblock [Online]. Available: \url{https://media-www.micron.com/-/media/client/global/documents/products/data-sheet/dram/gddr/gddr6/gddr6_sgram_8gb_brief.pdf}.

\bibitem{norrie2021design}
Thomas Norrie, Nishant Patil, Doe~Hyun Yoon, George Kurian, Sheng Li, James Laudon, Cliff Young, Norman Jouppi, and David Patterson.
\newblock The design process for google's training chips: Tpuv2 and tpuv3.
\newblock {\em IEEE Micro}, 41(2):56--63, 2021.

\bibitem{nvidia2021a100}
NVIDIA.
\newblock Nvidia a100 tensor core gpu.
\newblock [Online]. Available: \url{https://www.nvidia.com/en-us/data-center/a100/}.

\bibitem{openai}
OpenAI.
\newblock Input:output token ratio.
\newblock [Online]. Available: \url{https://beta.openai.com/docs/usage-guidelines/use-case-guidelines}.

\bibitem{radford2019language}
Alec Radford, Jeffrey Wu, Rewon Child, David Luan, Dario Amodei, and Ilya Sutskever.
\newblock Language models are unsupervised multitask learners.
\newblock {\em OpenAI blog}, 1(8):9, 2019.

\bibitem{rixner2000memory}
Scott Rixner, William~J Dally, Ujval~J Kapasi, Peter Mattson, and John~D Owens.
\newblock Memory access scheduling.
\newblock {\em ACM SIGARCH Computer Architecture News}, 28(2):128--138, 2000.

\bibitem{x330}
SAPEON.
\newblock {Product of SAPEON - X330}.
\newblock [Online]. Available: \url{https://www.sapeon.com/products/sapeon-x330}.

\bibitem{shin2018mcdram}
Hyunsung Shin, Dongyoung Kim, Eunhyeok Park, Sungho Park, Yongsik Park, and Sungjoo Yoo.
\newblock Mcdram: Low latency and energy-efficient matrix computations in dram.
\newblock {\em IEEE Transactions on Computer-Aided Design of Integrated Circuits and Systems}, 37(11):2613--2622, 2018.

\bibitem{shoeybi2019megatron}
Mohammad Shoeybi, Mostofa Patwary, Raul Puri, Patrick LeGresley, Jared Casper, and Bryan Catanzaro.
\newblock Megatron-lm: Training multi-billion parameter language models using model parallelism.
\newblock {\em arXiv preprint arXiv:1909.08053}, 2019.

\bibitem{vaswani2017attention}
Ashish Vaswani, Noam Shazeer, Niki Parmar, Jakob Uszkoreit, Llion Jones, Aidan~N Gomez, {\L}ukasz Kaiser, and Illia Polosukhin.
\newblock Attention is all you need.
\newblock {\em Advances in neural information processing systems}, 30, 2017.

\bibitem{wang2021spatten}
Hanrui Wang, Zhekai Zhang, and Song Han.
\newblock Spatten: Efficient sparse attention architecture with cascade token and head pruning.
\newblock In {\em 2021 IEEE International Symposium on High-Performance Computer Architecture (HPCA)}, pages 97--110. IEEE, 2021.

\bibitem{wang2019bfloat16}
Shibo Wang and Pankaj Kanwar.
\newblock Bfloat16: The secret to high performance on cloud tpus.
\newblock {\em Google Cloud Blog}, 4, 2019.

\bibitem{wolf2019huggingface}
Thomas Wolf, Lysandre Debut, Victor Sanh, Julien Chaumond, Clement Delangue, Anthony Moi, Pierric Cistac, Tim Rault, Rémi Louf, Morgan Funtowicz, Joe Davison, Sam Shleifer, Patrick von Platen, Clara Ma, Yacine Jernite, Julien Plu, Canwen Xu, Teven~Le Scao, Sylvain Gugger, Mariama Drame, Quentin Lhoest, and Alexander~M. Rush.
\newblock Huggingface's transformers: State-of-the-art natural language processing.
\newblock {\em arXiv preprint arXiv:1910.03771}, 2019.

\bibitem{vcu118}
Xilinx.
\newblock {Xilinx VCU118 Evaluation Kit}.
\newblock [Online]. Available: \url{https://www.xilinx.com/products/boards-and-kits/vcu118.html}.

\bibitem{SparseMICRO22}
Amir Yazdanbakhsh, Ashkan Moradifirouzabadi, Zheng Li, and Mingu Kang.
\newblock Sparse attention acceleration with synergistic in-memory pruning and on-chip recomputation.
\newblock In {\em 55th {IEEE/ACM} International Symposium on Microarchitecture, {MICRO} 2022, Chicago, IL, USA, October 1-5, 2022}, pages 744--762. {IEEE}, 2022.

\bibitem{yu2022nn}
Joonsang Yu, Junki Park, Seongmin Park, Minsoo Kim, Sihwa Lee, Dong~Hyun Lee, and Jungwook Choi.
\newblock Nn-lut: neural approximation of non-linear operations for efficient transformer inference.
\newblock In {\em Proceedings of the 59th ACM/IEEE Design Automation Conference}, pages 577--582, 2022.

\bibitem{zhou2022transpim}
Minxuan Zhou, Weihong Xu, Jaeyoung Kang, and Tajana Rosing.
\newblock Transpim: A memory-based acceleration via software-hardware co-design for transformer.
\newblock In {\em 2022 IEEE International Symposium on High-Performance Computer Architecture (HPCA)}, pages 1071--1085. IEEE, 2022.

\end{thebibliography}

\end{document}